\documentclass[prd,twocolumn,nofootinbib,notitlepage,aps,tightenlines,
preprintnumbers,amsmath,amssymb,amsfonts,showpacs,superscriptaddress]{revtex4-1}

\RequirePackage[colorlinks=true
,urlcolor=blue
,anchorcolor=blue
,citecolor=blue
,filecolor=blue
,linkcolor=blue
,menucolor=blue
,linktocpage=true
,pdfproducer=medialab
,pdfa=true
]{hyperref}

\usepackage{graphicx,epstopdf}
\usepackage{mathtools}
\usepackage{natbib,bm}
\usepackage[usenames,dvipsnames]{color}
\usepackage{slashed}    
\usepackage{hyperref}
\usepackage{multirow}
\usepackage{grffile}

\usepackage{float}
\usepackage{mathtools, nccmath}
\usepackage{lipsum}
\usepackage{cancel}

\def\be{\begin{equation}}
\def\ee{\end{equation}}
\def\ba{\begin{eqnarray}}
\def\ea{\end{eqnarray}}
\def\ge{\mathrel{\raise.3ex\hbox{$>$\kern-.75em\lower1ex\hbox{$\sim$}}}}
\def\la{\mathrel{\raise.3ex\hbox{$<$\kern-.75em\lower1ex\hbox{$\sim$}}}}

\def\thesection{\arabic{section}}
\def\theequation{\arabic{equation}}
\def\simgt{\mathrel{\raise.3ex\hbox{$>$\kern-.75em\lower1ex\hbox{$\sim$}}}}
\def\simlt{\mathrel{\raise.3ex\hbox{$<$\kern-.75em\lower1ex\hbox{$\sim$}}}}

\newcommand{\nc}{\newcommand}

\nc{\gone}{\bar g_{\pi NN}^{(1)}}
\nc{\gzero}{\bar g_{\pi NN}^{(0)}}
\nc{\al}{\alpha}
\nc{\ga}{\gamma}
\nc{\de}{\delta}
\nc{\ep}{\epsilon}
\nc{\ze}{\zeta}
\nc{\et}{\eta}
\nc{\ka}{\kappa}
\nc{\rh}{\rho}
\nc{\si}{\sigma}
\nc{\ta}{\tau}
\nc{\up}{\upsilon}
\nc{\ph}{\phi}
\nc{\ch}{\chi}
\nc{\ps}{\psi}
\nc{\om}{\omega}
\nc{\Ga}{\Gamma}
\nc{\De}{\Delta}
\nc{\La}{\Lambda}
\nc{\Si}{\Sigma}
\nc{\Up}{\Upsilon}
\nc{\Ph}{\Phi}
\nc{\Ps}{\Psi}
\nc{\Om}{\Omega}
\nc{\ptl}{\partial}
\nc{\del}{\nabla}
\nc{\ov}{\overline}
\nc{\newcaption}[1]{\centerline{\parbox{15cm}{\caption{#1}}}}
\nc{\us}{U(1)$_S$}

\def\beq{\begin{equation}}
\def\eeq{\end{equation}}
\def\bmat{\begin{displaymath}}
\def\emat{\end{displaymath}}
\def\bear{\begin{eqnarray}}
\def\eear{\end{eqnarray}}
\def\ba{\begin{eqnarray}}
\def\ea{\end{eqnarray}}
\def\bery{\begin{array}}
\def\ery{\end{array}}
\def\bit{\begin{itemize}}
\def\eit{\end{itemize}}
\def\ben{\begin{enumerate}}
\def\een{\end{enumerate}}
\def\btab{\begin{tabular}}
\def\etab{\end{tabular}}
\def\btbl{\begin{table}}
\def\etbl{\end{table}}
\def\bfig{\begin{figure}[htb]}
\def\efig{\end{figure}}
\def\bpic{\begin{picture}}
\def\epic{\end{picture}}

\def\nnl{\nonumber \\}

\def\ga{\mathrel{\raise.3ex\hbox{$>$\kern-.75em\lower1ex\hbox{$\sim$}}}}
\def\la{\mathrel{\raise.3ex\hbox{$<$\kern-.75em\lower1ex\hbox{$\sim$}}}}
\def\gappeq{\mathrel{\rlap {\raise.5ex\hbox{$>$}}
{\lower.5ex\hbox{$\sim$}}}}
\def\lappeq{\mathrel{\rlap{\raise.5ex\hbox{$<$}}
{\lower.5ex\hbox{$\sim$}}}}

\def\gyr{{\rm \, G\kern-0.125em yr}}
\def\mev{{\rm \, Me\kern-0.125em V}}
\def\gev{{\rm \, Ge\kern-0.125em V}}
\def\tev{{\rm \, Te\kern-0.125em V}}

\begin{document}

\title{Dark Sector Production via Proton Bremsstrahlung}

\author{Saeid Foroughi-Abari}
\affiliation{Department of Physics and Astronomy, University of Victoria, 
Victoria, BC V8P 5C2, Canada}

\author{Adam Ritz}
\affiliation{Department of Physics and Astronomy, University of Victoria, 
Victoria, BC V8P 5C2, Canada}

\date{August 2021}

\begin{abstract}
\noindent 

Experiments using proton beams at high luminosity colliders and fixed target facilities provide impressive sensitivity to new light weakly coupled degrees of freedom. With these experiments in mind, we revisit the production of dark vectors and scalars via proton bremsstrahlung, making use of a model that describes the underlying nucleon scattering cross-section in the forward direction due to pomeron exchange. We compare the resulting distributions and rates with those obtained via variants of the Fermi-Weizsacker-Williams approximation, and provide production rate distributions for a range of beam energies, including those relevant for the proposed Forward Physics Facility at the High Luminosity-LHC.

\end{abstract}
\maketitle

\section{Introduction}

The strongest empirical evidence for new physics beyond the Standard Model (SM), in particular for dark matter and neutrino mass, may hint at the presence of a more complex dark sector \cite{pospelov2008,Batell:2009yf,Essig:2009nc,Reece:2009un,Freytsis:2009bh,Batell:2009jf,Freytsis:2009ct,Essig:2010xa,Essig:2010gu,McDonald:2010fe,Williams:2011qb,Abrahamyan:2011gv,Archilli:2011zc,Lees:2012ra,Davoudiasl:2012ag,Kahn:2012br,Andreas:2012mt}. Such scenarios necessarily imply the presence of new degrees of freedom which are weakly coupled to the SM, and could therefore be light relative to the weak scale. This framework has been studied in great detail in recent years (see e.g. \cite{DS16,CV17,PBC}), as there is sensitivity to light weakly-coupled degrees of freedom at a variety of luminosity frontier experiments, including proton \cite{Batell:2009di,deNiverville:2011it,deNiverville:2012ij,Kahn:2014sra,Adams:2013qkq,Soper:2014ska,Dobrescu:2014ita,Coloma:2015pih,dNCPR,MB1,MB2,Feng:2017uoz,Alpigiani:2018fgd,Ariga:2018pin,Dutta:2020vop,Batell:2021blf,Batell:2021aja} and electron \cite{Bjorken:2009mm,Izaguirre:2013uxa,Diamond:2013oda,Izaguirre:2014dua,Batell:2014mga,Lees:2017lec,Berlin:2018bsc,NA64:2019imj,Berlin:2020uwy} fixed target facilities and colliders. 

The dark sector framework relies on minimal assumptions, but effective field theory provides a simplifying perspective that helps to classify the interactions of new neutral states with the Standard Model (SM) according to their dimensionality. There are only three relevant or marginal `portal' operators that are unsuppressed by a new (potentially high) energy scale. These Higgs, vector and neutrino portals therefore comprise a priori the leading couplings of the SM to a dark or hidden sector. Significant theoretical and experimental effort has been invested in studying these portal interactions, motivated in part by their importance for the phenomenology of light dark matter (DM) models  \cite{DS16,CV17,PBC}. The most relevant production channels for dark force mediators are therefore of importance for associated searches at collider and fixed target facilities. For experiments making use of proton beams, the dominant production channels depend on the energy of the proton beam and the mass of the dark mediator. Among them, bremsstrahlung of dark vectors and scalars can be particularly important over the dark sector mass range from about 500 MeV to a GeV. However, computing the bremsstrahlung production rate, particularly in the forward direction, is difficult as it involves the nonperturbative physics of the forward $pp$ (or $pn$) cross section. Thus far, most analyses have relied on variants of the Weizsacker-Williams (WW) approximation, developed in the 1970's as a generalization of the successful approach used for electron beams. 

\begin{figure*}[t]
    \centering
    \includegraphics[width=0.49\textwidth]{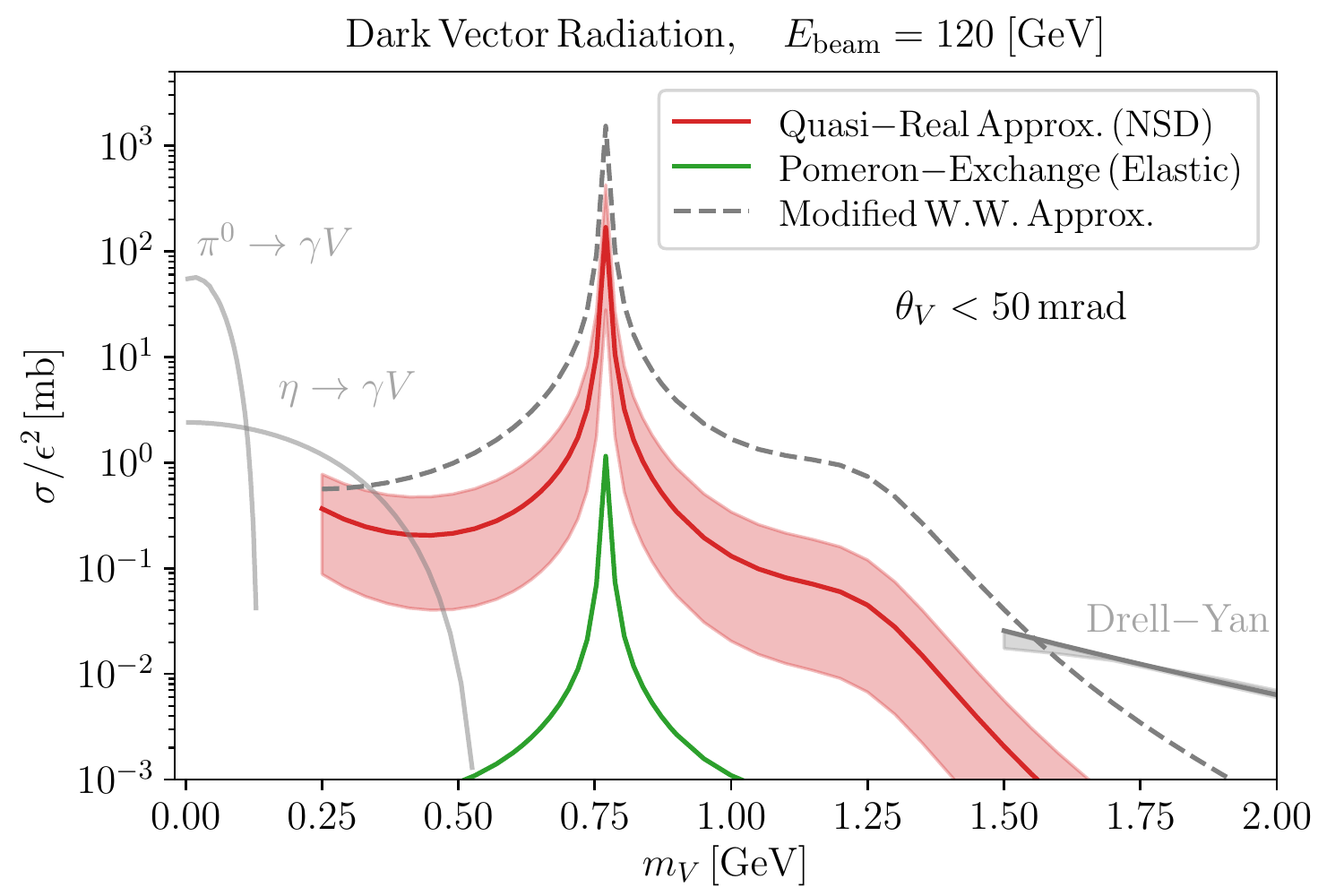} 
    \includegraphics[width=0.49\textwidth]{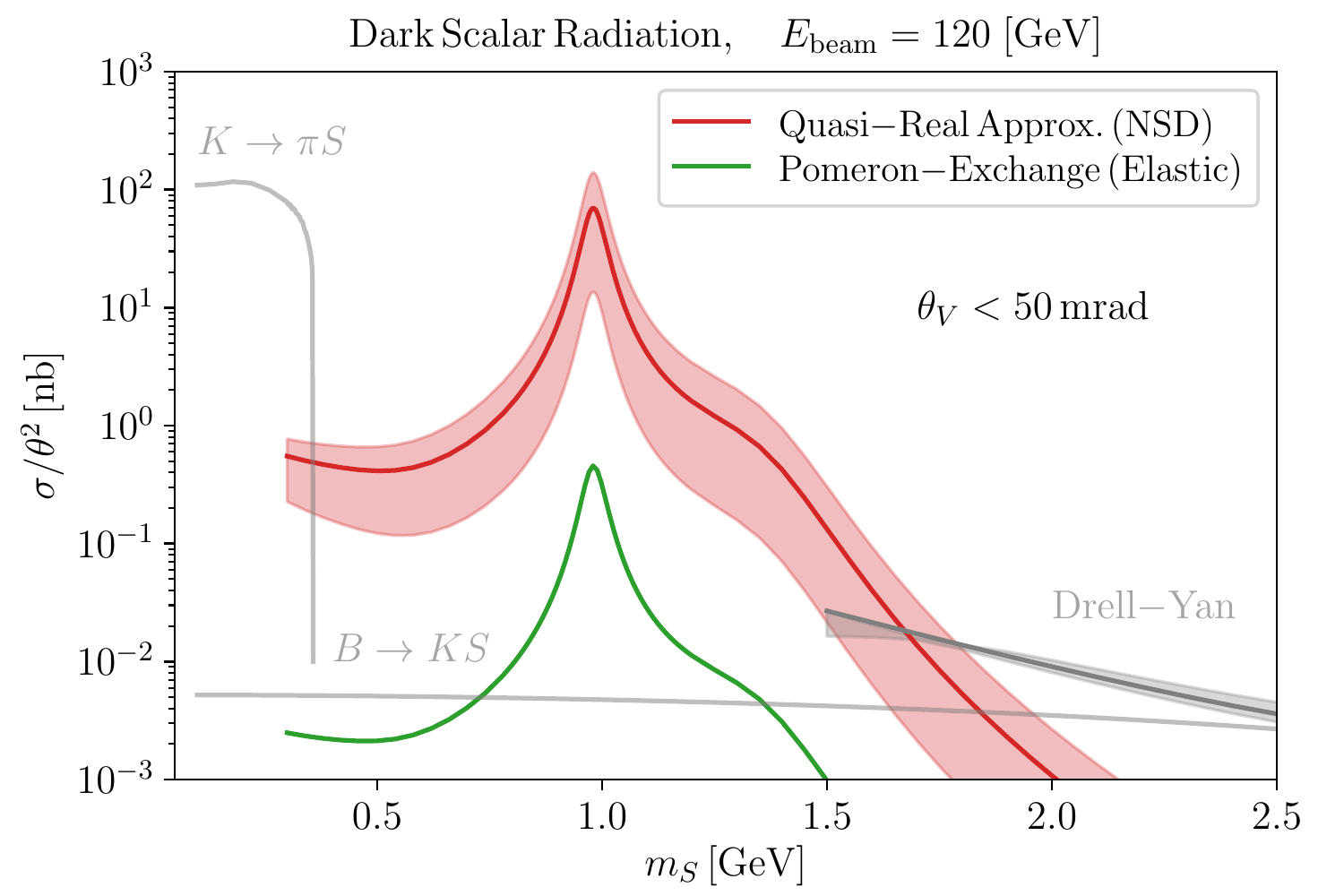}
    \caption{The production cross section of dark vectors and scalars for a $120 \gev$ fixed target beam as a function of mass and within $\theta < 50$~mrad of the beam axis (lab frame). The red curves denote the rates using the quasi-real approximation in non-single diffractive scattering, and the uncertainty band corresponds to varying the associated cut-off scale $\Lambda_p\in [1,2]~\gev$ with the central value $1.5~\gev$. The green curves show the associated rates from initial and final state radiation in quasi-elastic scattering, where interference effects cause a significant suppression. In the vector case, the dashed grey curve uses the modified WW approximation of \cite{Blumlein:2013cua} with a cut on transverse momentum $p_T<1~\gev$, while for both plots the lighter grey curves show other production channels from meson decay~\cite{Batell:2009di,dNCPR}, and parton-level Drell-Yan~\cite{deNiverville:2012ij} processes relevant at higher mass. See the text in Sections 3 and 4 for further details.}
    \label{fig:Production120GeV}
\end{figure*}

\begin{figure*}[t]
    \centering
    \includegraphics[width=0.49\textwidth]{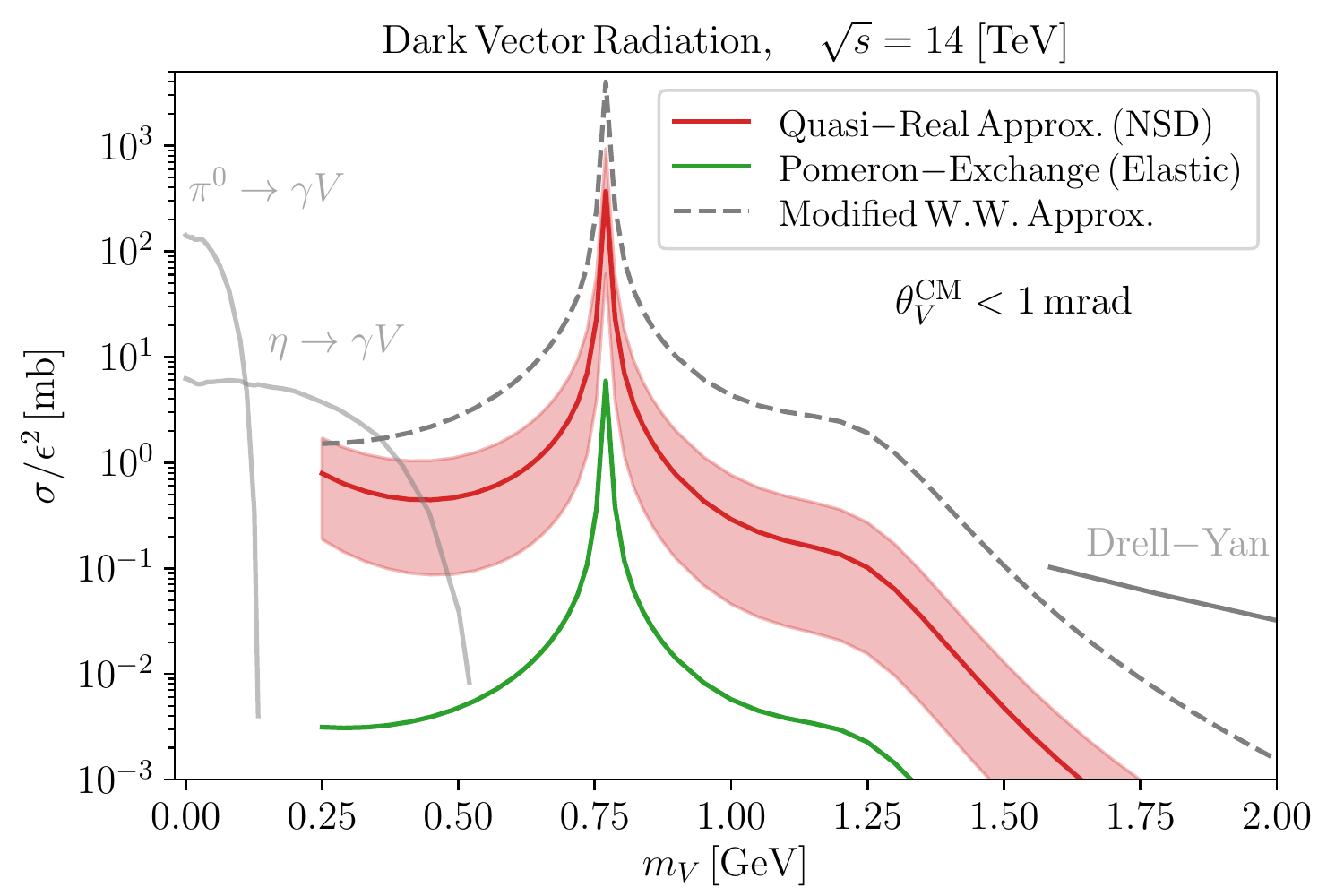}
    \includegraphics[width=0.49\textwidth]{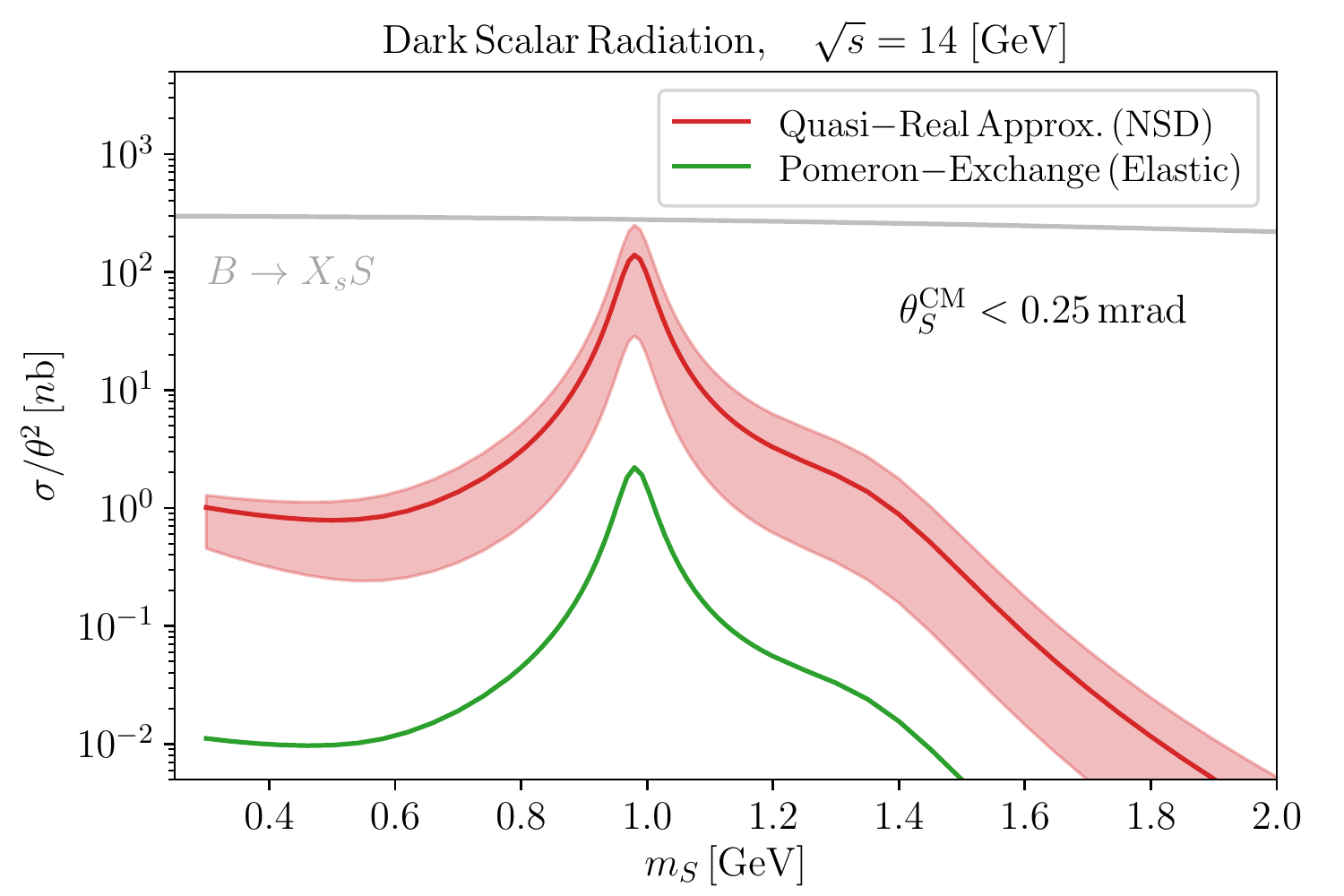} 
    \caption{The production cross section of dark vectors and scalars at $14 \tev$ (centre of mass) energy as a function of mass and within $\theta < 1$ or $0.25$~mrad of the beam axis (centre of mass frame). The curves are as described in Fig.~\ref{fig:Production120GeV}. The lighter grey curves again show other production channels from meson decay~\cite{Feng:2017uoz,Kling:2021fwx}, and parton-level Drell-Yan~\cite{Kling:2021fwx} processes relevant at higher mass. See the text in Sections 3 and 4 for further details.}
    \label{fig:Production14TeV}
\end{figure*}

In this paper, we revisit the production of dark vectors and scalars via proton bremsstrahlung and build a model to describe this process in the context of pomeron-mediated forward scattering. We use this model to analyze the various production modes associated with initial and final state radiation in diffractive and non-diffractive proton scattering. We will focus our attention on proton-proton scattering, as a generic contribution relevant for both the High Luminosity-LHC (HL-LHC) and for fixed target experiments. This analysis will allow us to test the impact of various approximations and kinematic constraints, and to compare this approach with the modified WW approximation~\cite{Blumlein:2013cua}. Our final results for production rates are shown in Figure~\ref{fig:Production120GeV} for a 120 GeV fixed target beam at Fermilab, and Figure~\ref{fig:Production14TeV} for the 14 TeV LHC, along with various comparisons.

The rest of this paper is organized as follows. In the next Section, we provide a brief overview of dark sector production channels in proton beam experiments, and then turn in Section 3 to a discussion of bremsstrahlung in forward proton scattering. We discuss the modeling of forward elastic and diffractive scattering via pomeron exchange, and then build a model for initial and final state radiation of dark vectors and scalars via this process, along with a more generic model for initial state radiation in non-single-diffractive scattering. We present our results in Section 4, along with comparisons to modified WW approaches, and conclude in Section 5. A number of technical details are relegated to a series of appendices.

\section{Overview of dark sector production via proton beams}

We will consider dark sectors which couple to the Standard Model via vector and/or scalar portal couplings of the form,
\be
 {\cal L} \supset - \frac{1}{2} \ep F^{\mu\nu} F'_{\mu\nu} - ASH^\dagger H + \cdots \label{portal}
\ee
where $F'_{\mu\nu} = \ptl_\mu A'_\nu - \ptl_\nu A'_\mu$, which induces couplings of the dark vector $A'_\mu$ to the electromagnetic current, and the dark scalar $S$ to the scalar current which contains all scalar bilinears of charged fermions.

In a proton collider or fixed target experiment, there are a variety of production modes, which proceed via an on or off-shell mediator $A'_\mu$ or $S$. For the low sub-GeV mass range of interest, meson decays provide an important channel. For the vector portal, the decays $\pi^0/\et \rightarrow \gamma + A'^*$, with subsequent visible or invisible decays of $A'^*$ provide the dominant channel for $m_{A'} < 0.5$ GeV \cite{Batell:2009di,dNCPR}, while for scalars, the decays $K\rightarrow \pi + S^*$ and $B \rightarrow K + S^*$ provide the dominant channels for a larger mass range if the beam is sufficiently energetic \cite{Batell:2009jf}. For proton beam fixed target experiments, secondary electromagnetic bremsstrahlung may also be important for very low masses \cite{Celentano:2020vtu}. On the other hand, for mediator masses well above the nucleon scale, Drell-Yan processes such as $q\bar{q} \rightarrow A'^*$ or loop-induced scalar production $gg \rightarrow S^*$ are relevant \cite{deNiverville:2012ij}. 

In the present paper, our focus is on the intermediate mass range $m_{A'/S} \sim $ GeV, where mixing of the dark states with vector or scalar mesons with the same quantum numbers can resonantly enhance the production rate. In the case of forward production, where the small sub-GeV momentum transfer allows us to treat the coupling of the dark sector state with the proton collectively, this channel amounts to `dark bremsstrahlung'. For the scalar and vector portals, the induced coupling to protons, which will be relevant here, follows directly from (\ref{portal}),
\be 
 {\cal L}_{\rm eff} \supset - \ep e A'_\mu \bar{p} \gamma^\mu p  - g_{SNN} \theta S \bar{p} p +\cdots
\ee
where we have ignored higher multipole couplings for $A'_\mu$, the $h-S$ mixing angle $\theta \simeq Av/m_h^2 \ll 1$ for the parameters of interest, and $g_{SNN} = 1.2 \times 10^{-3}$ can be obtained via the use of low energy theorems \cite{Boiarska:2019jym} (see also \cite{Alarcon:2011zs,Alarcon:2012nr}). In the next section, we will consider a model for dark bremsstrahlung based on these couplings, and the underlying physics of proton-proton scattering.

\section{Proton Bremsstrahlung}

We will focus our attention in this paper on bremsstrahlung, which is important for the forward production of dark sector mediators with hadronic scale mass, particularly due to the possibility of resonant mixing with hadronic states. This is a complex process to model for proton beams, and we will consider several different approximation strategies, which will allow an assessment of relative precision. In particular, we will compare four different approaches:
\begin{itemize}
\item ISR and FSR in quasi-elastic scattering
\item ISR in non-single diffractive scattering via the quasi-real approximation
\item Hadronic generalization of the WW approximation
\item Modified WW approximation
\end{itemize}
These approaches are described in more detail below, with some technical details relegated to Appendices. In all cases, a timelike form-factor for coupling to the proton provides resonant enhancements, and is discussed separately.

\subsection{Modeling forward \texorpdfstring{$pp$}{Lg} scattering}
We start by reviewing the high-energy behaviour of hadronic scattering processes with small momentum transfer that cannot be described in terms of perturbative QCD. In high energy $pp$ collisions, where soft interactions play a dominant role, the total cross-section can be divided into diffractive and non-diffractive scattering processes \cite{Zyla:2020zbs,Frankfurt:2013ria}. In elastic diffractive scattering both protons stay intact after the collision while in inelastic diffractive scattering, one of the incoming protons or both dissociate into multi-particle final states with the invariant mass $M\ll \sqrt{s}$, preserving the quantum number(s) of the associated initial proton(s). Non-diffractive scattering denotes more generic inelastic processes, and is the characteristic process used at the LHC to observe new physics events with large transverse momentum. Our focus here is instead on the forward region, and processes with GeV-scale or sub-GeV momentum transfer.

\begin{figure}[t]
    \centering
    \includegraphics[width=0.4\textwidth]{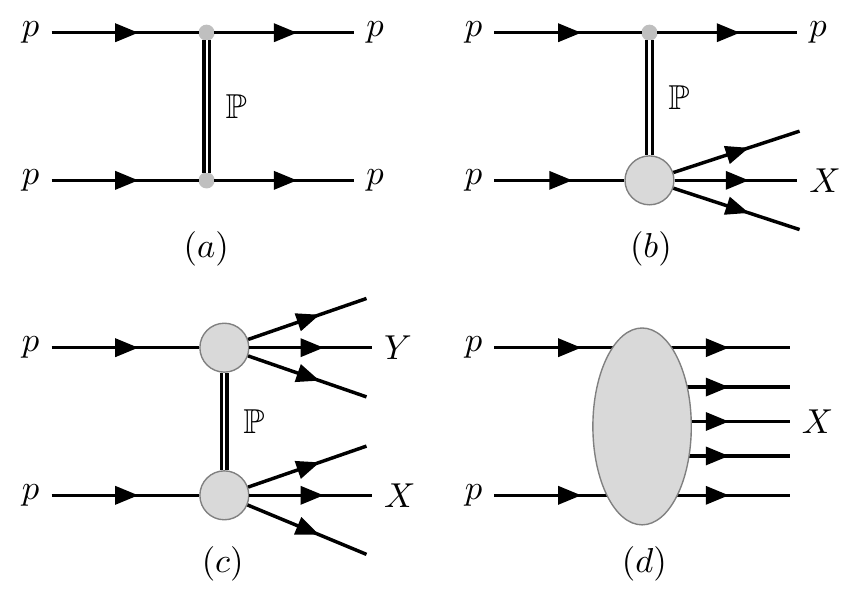}\caption{Schematic diagrams for the lowest order Pomeron exchange processes contributing to (a) elastic scattering, (b) single dissociation, (c) double dissociation and (d) non-diffractive interactions. The double line $\mathbb{P}$ corresponds to the Pomeron exchange and $p$ for proton.}
    \label{fig:Diffractive}
\end{figure}

Within the category of diffractive scattering, single dissociation (SD), corresponding to $pp{\rightarrow} p{+}X$, and double dissociation (DD), corresponding to $pp{\rightarrow} X{+}Y$, have the following characteristics: \textit{i}) the diffracted state is separated from the scattered proton by a large rapidity gap devoid of any hadronic activity; \textit{ii}) the energy transfer between the two interacting protons remains small; and \textit{iii}) the coherence condition implies $\xi {=}\ M_{X}^2/s \lesssim 0.15$ which separates dissociation from the inelastic process. Such processes have traditionally been modeled  phenomenologically with Regge exchanges, along with single or multi-pomeron exchange. Feynman diagrams corresponding to one pomeron exchange in elastic, single- and double-diffraction processes are shown in Fig.~\ref{fig:Diffractive}, where the remaining configurations correspond to non-diffractive interactions. Experimental data indicates that the high-energy total and elastic $pp$ cross sections grow slowly with centre of mass energy, and have the asymptotic behavior $\sigma_{\rm tot} \sim \ln(s)^2$ \cite{Donnachie:1992ny}. At LHC energies, diffractive processes constitute up to $40\%$ of the total $pp$ cross section \cite{Abelev:2012sea}. 

The Donnachie-Landshoff (DL) model of diffractive $pp$ scattering incorporates the Regge theory approach which sums the exchanges of many particles and provides a good description of the existing elastic differential cross section data, including the exponential fall at low Mandelstam $t$, the dip region at mid $t$, and the rapid fall in cross section at high $t$. The DL parameterization utilizes single Regge (with $\rho, \omega$ and $f, a_2$ trajectories) and Pomeron exchange \cite{Donnachie:1983ff}, along with multiple Regge and Pomeron exchanges \cite{Donnachie:1984xq,Donnachie:2011aa,Donnachie:2013xia}, plus triple-gluon exchange for $|t| \gtrsim 3.5 \mathrm{GeV}^2$ \cite{Donnachie:1996rq}. This parametrization will form the basis of our bremsstrahlung model, and we review the components in more detail below.

\begin{figure}[t]
    \centering
    \includegraphics[width=0.49\textwidth]{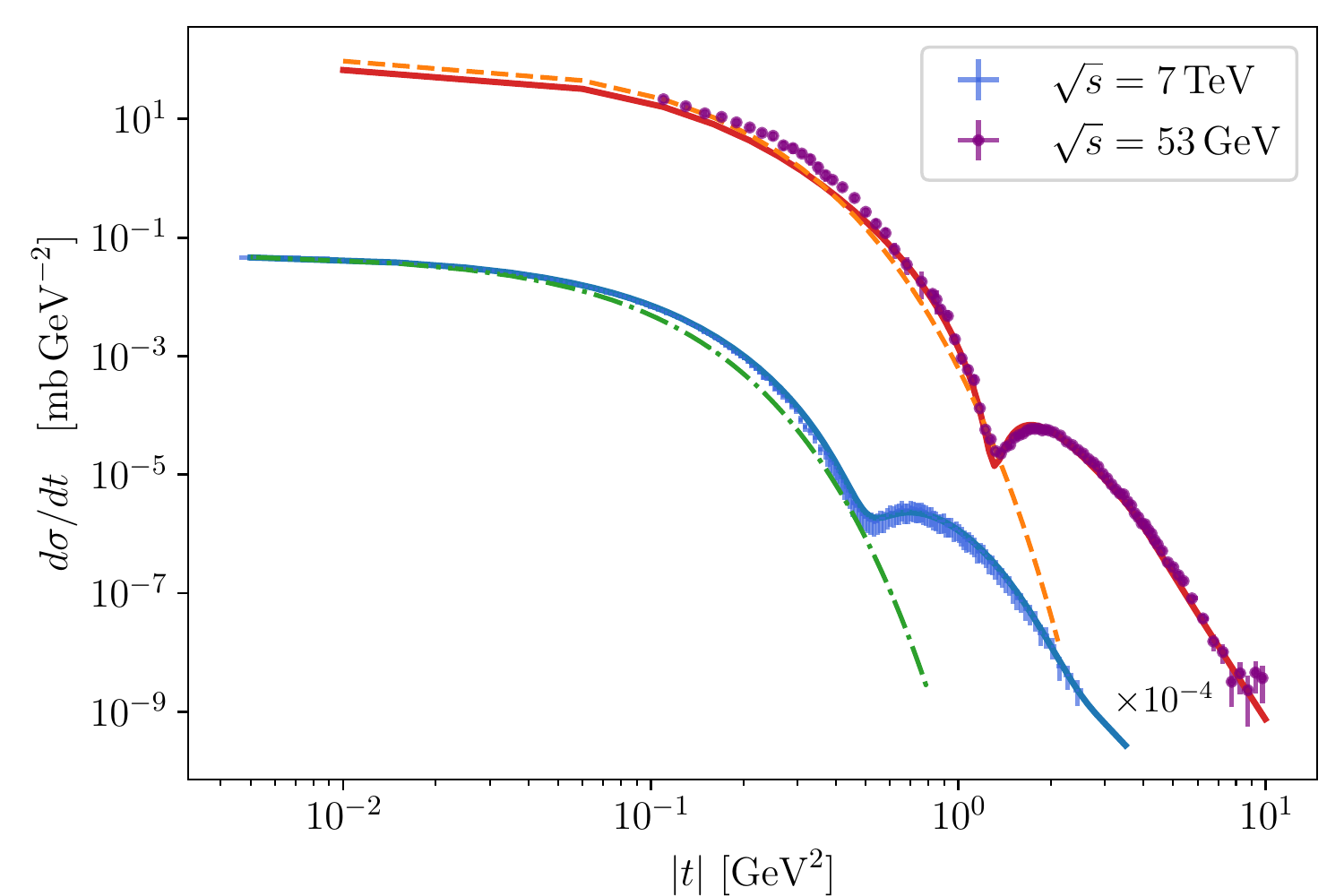}
    \caption{The DL model fit for the elastic differential cross-section compared with $pp$ data at $\sqrt{s}=7\, \mathrm{TeV}$ \cite{TOTEM:2013lle,ATLAS:2014vxr} (in blue) and $\sqrt{s}=53\, \mathrm{GeV}$ \cite{Breakstone:1985pe} (in purple). A single soft pomeron exchange fit is also shown in each case for small $t$-values with dashed lines. The diffractive dip requires the addition of double pomeron exchange and other components of the full model.}
    \label{fig:Elastic}
\end{figure}

\textit{Elastic scattering:-} We first review the elastic cross-section. The full parametrization of multi-Regge and pomeron exchanges in the DL model is summarized in Appendix A, and we present a comparison of this model to data in Fig.~\ref{fig:Elastic}. It is notable that just including single-pomeron exchange describes the measured $pp$ elastic cross section at high-energies remarkably well for sufficiently small $t$. For later purposes, we find it convenient to present this parametrization in terms of a phenomenological soft pomeron propagator $G_{\mathbb{P}}(s,t)g^{\mu\nu}$ and an effective proton-pomeron vertex $\Gamma_{\mathbb{P}}^{\mu}(t)$,
\begin{equation} \label{propagator_single}
G_{\mathbb{P}}(s,t) = \frac{(2\nu\alpha_{\mathbb{P}}^{\prime})^{\alpha_{\mathbb{P}}(t)}}{2\nu}\eta_{\mathbb{P}}(t), \quad \Gamma^{\mu}(t)=-iY_{\mathbb{P}} F_{\mathbb{P}}(t) \gamma^{\mu},
\end{equation}
where $2\nu = (s-u)/2$. The effective soft pomeron trajectory is linear in $t$,
\begin{equation}
\alpha_{\mathbb{P}}(t)=1+\epsilon_{\mathbb{P}}+\alpha_{\mathbb{P}}^{\prime}t,
\end{equation}
where the intercept $\alpha_{\mathbb{P}}(0)>1$,  
and $Y_{\mathbb{P}}$ is the coupling strength of the pomeron to the proton. The parameter values in these fits are provided in Appendix A. The pomeron form factor was traditionally assumed to have a dipole form~\cite{Donnachie:1983ff}, $F_{\mathbb{P}}(t)\sim 1/(1-t/0.71\, {\rm GeV}^2)^2$, as for the proton electromagnetic form factor. However, more recent studies~\cite{Donnachie:2013xia} utilize an exponential form factor, $F_{\mathbb{P}}^2(t) = A\exp(at){+} (1{-}A)\exp(bt)$. Finally, $\eta_{\mathbb{P}}(t) {=}-\exp{(-\frac{1}{2}i\pi\alpha_{\mathbb{P}}(t))}$ is the signature factor. 

As is apparent in Fig.~\ref{fig:Elastic}, the cross section modeled with soft pomeron exchange in the region where the squared momentum transfer $t$ is not too large can be approximated by a simple exponential fall-off $d\sigma/dt \propto e^{-B|t|}$. Note that with increasing energy the differential cross section becomes steeper and the diffractive slope $B$ which grows linearly in $\log(s)$ (the so called shrinkage of the diffractive peak) has been measured by several experiments \cite{Antchev:2017dia} and is $\sim 20 \gev^{-2}$ at LHC energies. 

While single pomeron exchange is sufficient to model the elastic cross section for small $t$, the inclusion of higher exchanges, including double pomeron exchange, becomes important for fitting the diffractive dip apparent in Fig.~\ref{fig:Elastic} for $|t|\gtrsim1$ GeV. These additional components of the model are described in Appendix A. 
Considering all the contributions from single-pomeron ($\mathbb{P}$) and double-pomeron ($\mathbb{PP}$) exchange, and triple-gluon ($3g$) exchange, the elastic $pp$-scattering at high energies takes the form
\begin{equation} \label{SigmaEl}
\frac{d \sigma^{\rm el}}{dt} \simeq \frac{1}{4\pi}|\mathcal{A}_{\rm el}|^2,
\end{equation}
where
\begin{equation} \label{PomeronAmp}
\mathcal{A}_{\rm el}(s,t) = \sum_{\mathbb{P},\mathbb{PP}}(Y_{i} F_{i}(t))^2 G_{i}(s,t) + Y_{3g}^2G_{3g}(t).
\end{equation}

\textit{Dissociative scattering:-} Scattering with single diffractive (SD) dissociation of one proton can be modeled with the triple-pomeron formalism using a generalized optical theorem \cite{Khoze:2000wk,Khoze:2014aca}, in which the corresponding $pp{\rightarrow} p{+}X$ cross section is given by
\begin{align} \label{SigmaSD}
\frac{d \sigma^{\rm SD}}{dtdM_{X}^2} &= \frac{g_{3\mathbb{P}}(t)}{16\pi^2}\frac{g_{\mathbb{P}}(0)g_{\mathbb{P}}(t)^2}{M_{X}^2} \\
 & \qquad \qquad \times\bigg(\frac{s}{M_{X}^2}\bigg)^{2\alpha_{\mathbb{P}}(t){-}2}\bigg(\frac{M_{X}^2}{s_0}\bigg)^{\alpha_{\mathbb{P}}(0){-}1}, \nonumber
\end{align}
where the diffractive mass is $M_{X}^2=\xi s$, with $\xi\lesssim 0.15$, $g_{\mathbb{P}}(t)$ is the \textit{soft} pomeron-proton coupling strength with an exponential $t$-dependent form-factor, and $g_{3\mathbb{P}}(t)$ is the triple-pomeron coupling. The dimensionful coupling $g_{\mathbb{P}}^2(0)\approx 57$ mb specifies $\sigma_{\rm tot}$, and is distinct from the single pomeron-exchange value defined previously, while $g_{3\mathbb{P}}(0)/g_\mathbb{P}(0)\simeq 0.2$ is obtained from
a triple-Regge analysis of lower energy data \cite{Khoze:2014aca}. In the low-mass regime $M_{X} \lesssim t$, the system $X$ is dominated by baryon resonances and low level excited states of the proton \cite{Appleby:2016ask,Rasmussen:2018dgo}.

In the following subsections, where we consider initial and final state radiation for these exchange processes,  it will be useful to have parameterizations for the total $pp$ cross section (in mb) taken from experimental data \cite{PDG2016,PhysRevLett.89.201801},
\begin{equation}
\sigma_{\rm tot}(s) = 34.4 + 0.3\log^2(s/s_0)+13.1(\frac{s}{s_0})^{-\eta_1}+7.4(\frac{s}{s_0})^{-\eta_2},
\end{equation} 
where 
$s_0 =15.98 ~\gev^2$, $\eta_1=0.45$, $\eta_2=0.55$ and similarly the elastic scattering cross section (in mb) \cite{Antchev:2017dia} ,
\begin{equation}
\sigma_{\rm el}(s) = 11.8 - 1.6\log(s) + 0.14\log^2(s).
\end{equation} 

The single diffractive cross section can similarly be modeled \cite{Kaidalov:2009aw,Gotsman:2012rm} and parametrized based on the experimental data~\cite{Abelev:2012sea}. However, larger systematic uncertainties in the diffractive cross sections of about $5{-}10\%$ (depending on the energy) arise due to the fact that at high energies, defining (and selecting) purely diffractive events is problematic~\cite{Abelev:2012sea}. The $\sqrt{s}=14$ TeV LHC cross-section for SD scattering is $\sim 10$ mb, while for DD scattering it is $\sim 7$ mb \cite{ALICE:2012fjm}. We also introduce the inelastic, non-single diffractive (NSD) cross section, $\sigma_{\rm NSD} \equiv \sigma_{\rm tot}{-}\sigma_{\rm el}{-}\sigma_{\rm SD}$, which can be parametrized following Ref.~\cite{Likhoded:2010pc} as
\begin{equation}
\sigma_{\rm NSD}(s)  = 1.76+19.8 \big(\frac{s}{\gev^2} \big)^{0.057} \quad {\rm mb}.
\end{equation}

\subsection{ISR and FSR in quasi-elastic scattering}\label{sec:QusaiElastic}
\begin{figure}[t]
\centering
    \includegraphics[width=0.49\textwidth]{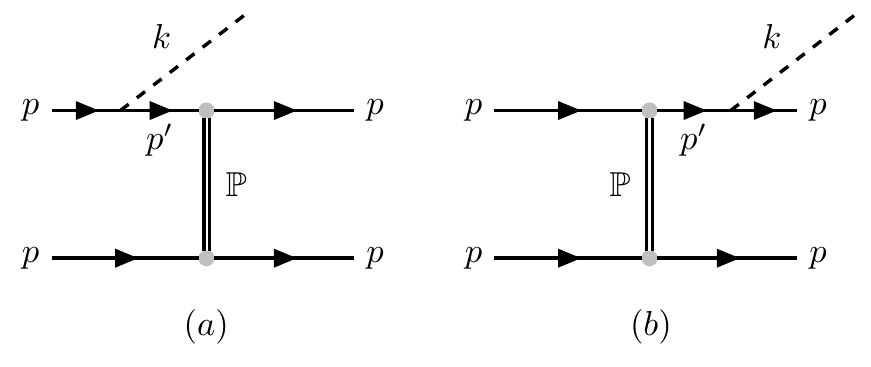}
    \caption{Dark state radiation from (a) initial state, and (b) final state proton bremsstrahlung through pomeron exchange. The label $p^{\prime}$ stands for the intermediate proton's momentum.}
    \label{fig:brem_el}
\end{figure}

We can build a model of proton bremsstrahlung by adding initial state radiation (ISR) and final state radiation (FSR) to the pomeron exchange model for $pp$ elastic scattering,  as represented by the two Feynman diagrams in Fig.~\ref{fig:brem_el}. Note that emission from the target proton into the forward region is negligible compared with that from the ultra-relativistic beam proton (as discussed for the photon bremsstrahlung in electron scattering \cite{BERESTETSKII1982354}). Also note that $u$-channel (exchange) diagrams are sub-leading in the regime of high energy but soft (small-$t$) scatterings, since $t\ll u\sim s$. 

Utilizing the phenomenological pomeron propagator and vertices outlined above, we compute the dark bremsstrahlung rate associated with the diagrams in Fig.~\ref{fig:brem_el}. Considering radiation from both incoming and outgoing beam protons with 4-momentum $k^{\mu}=(E_k,\vec{k})$, the contribution from the quasi-elastic process $pp\rightarrow ppD$ (where $D=V,S$) in the lab frame is given by
\begin{equation} \label{Brem_Elastic_CrossSection}
\frac{d^2 \sigma^{\rm el}_{pp\rightarrow ppD}}{dE_k d\cos \theta_k }{=}\frac{1}{64(2\pi)^4 p_p m_p^2}\frac{|\vec{k}|}{|\vec{p}_p{-}\vec{k}|}\int dt d\phi\, |\overline{\mathcal{M}^{pp\rightarrow ppD}}|^2 \, ,
\end{equation}
where $\theta_k$ is the dark vector/scalar emission angle with respect to the beam. The summed and averaged square of the 2 to 3 matrix element $ |\overline{\mathcal{M}^{pp\rightarrow ppD}}|^2$ is presented in the Appendix~\ref{app:MatEel}.

To compare with the other approximate methods of calculation, we will find it convenient to define the differential splitting probability of the proton to emit a dark state in the
form,
\begin{equation}\label{Split_Prob}
d\mathcal{P}^{\rm split.}_D = \frac{1}{\sigma^{\rm el}_{pp}(s)}\frac{d^2 \sigma^{\rm el}_{pp\rightarrow ppD}}{dE_k d\cos \theta_k }.
\end{equation}
The resulting splitting probability of dark state emission as a function of the scalar/vector energy $E_{k}$ and angle $\theta_k$ is shown in the next section in Figs.~\ref{fig:diffsplitVector} and \ref{fig:diffsplitScalar}, respectively. In the plots we also compare the complete ISR + FSR calculation via pomeron exchange with the result from ISR only, which as discussed below can be associated with NSD scattering with various final states. 
In considering radiation during quasi-elastic scattering, we observe a strong interference between the ISR and FSR amplitudes, and a significant cancelation which suppresses the final result as a generic feature of bremsstrahlung \cite{Lebiedowicz:2013xlb,Khoze:2010jv} for both scalar and vector cases. Similar results hold for varied choices of $m_D$, and emitted angles and energies. For completeness, we note that these radiative topologies are subject to soft and collinear divergences when the radiated particle is parametrically light, and certain divergences are only canceled on considering loop corrections to the underlying scattering process. We will not account for these effects here, as the mediator mass in the regimes of interest provides an infrared regulator that is sufficient to cut off those divergences.

Using the single diffractive cross section in Eq.~(\ref{SigmaSD}), one can also calculate the differential cross section for dark state radiation through $pp$ bremsstrahlung. This computation has not been done explicitly in this paper, but based on the large cancellation observed in the quasi-elastic regime, we anticipate a similar cancellation to also occur in the case of single diffractive topologies. In addition, since single diffractive events make up at most $10\%$ of the total cross-section at the relevant energies, we will focus now on the non-single diffractive topologies.

\subsection{ISR in non-single diffractive scattering and the quasi-real approximation}
\label{sec:OnShell}
In quasi-elastic and single diffractive scattering, radiation from both initial and final protons is tightly connected, and as observed above significant interference between the two amplitudes leads to a suppression of the total rate as compared to ISR or FSR alone. As a result, in order to identify the leading processes, we consider radiation in $pp$ non-single diffractive topologies, where both the beam proton and target proton dissociate after scattering. In such processes, radiation from particles other than proton in the final state should not interfere destructively with proton ISR, thus one expects no significant cancellation between ISR and FSR in non-single diffractive events. 

\begin{figure}[t]
    \centering
    \includegraphics[width=0.3\textwidth]{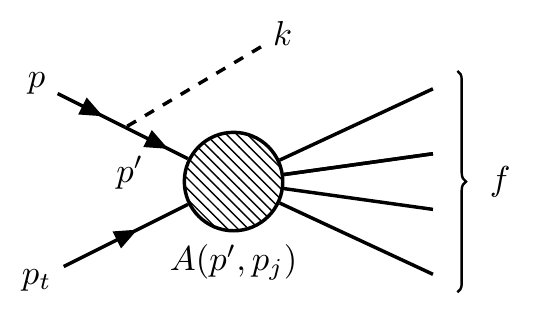}\caption{Dark sector initial state radiation in a generic non diffractive scattering event.}
    \label{fig:ISR}
\end{figure}

The ISR contribution can be estimated by artificially turning off the FSR amplitude in the consideration of quasi-elastic scattering above. However, to test this estimate, we now discuss another approach for evaluating the proton bremsstrahlung cross section, where instead of limiting the final state, the intermediate fermion propagator in the ISR diagram is approximated \cite{Kessler1960SurUM,Baier:1973ms,Baier:1980kx,Nicrosini:1988hw} within an on-shell approach (also known as time-ordered perturbation theory) used by Altarelli-Parisi \cite{Altarelli:1977zs}. We will refer to this as the `quasi-real approximation'.

We can compare the process of scattering of a beam proton by a target proton $pp_t\rightarrow X$, with another process, $pp_t \rightarrow X+D$ involving an additional dark state $D$ emitted from the incoming proton as shown in Fig.~\ref{fig:ISR}, along with any possible particles in the final state $X$. Under certain kinematic conditions, to be formulated below, the cross-section for the second process can be expressed in terms of the cross section of its sub-process, along with the splitting probability of emission of a single dark state in the collision.
Let us denote the corresponding amplitude for the hard scattering process without radiation as $\mathcal{M}_r^{pp_t\rightarrow f}=A(p,p_j)u^r(p)$,
where $u^r(p)$ is the spinor of the incoming proton with helicity $r$ and momentum $p$, and $A(p,p_j)$ is the remaining part of the amplitude for the hard scattering with $p_j$ denoting the momenta of the other particles in the process. The amplitude for dark ISR off the incoming proton with momentum $k$ can then be obtained from the amplitude for the original process by adding an external dark state line,
\begin{equation}\label{AmpEmission}
A(p,p_j) \rightarrow A(p-k,p_j)\frac{i(\cancel{p}-\cancel{k}+m_p)}{(p-k)^2-m_p^2}.
\end{equation} 
In the framework of the quasi-real approximation, which is best suited to the high energy limit, the intermediate proton propagator can be approximated as
\begin{equation}\label{PropApprox}
\frac{i(\cancel{p}-\cancel{k}+m_p)}{(p-k)^2-m_p^2} \approx \frac{i}{2E_{p^{\prime}}} \frac{\sum_{r^{\prime}}u^{r^{\prime}}(p-k)\bar{u}^{r^{\prime}}(p-k)}{E_p-E_k-E_{p^{\prime}}},
\end{equation}
where $E_p$, $E_k$, and $E_{p^{\prime}}=\sqrt{(\vec{p}-\vec{k})^2+m_p^2}$ are the energy of the incoming proton, the radiated dark state, and the intermediate proton, respectively.  Using this approximation, at the cost of being non-covariant, the numerator of the intermediate proton propagator in Eq.~(\ref{AmpEmission}) can be replaced by the polarization sum for an on-shell proton resulting in the following matrix element for dark state emission off the initial state proton,
\begin{equation}\label{}
\mathcal{M}_r^{pp_t\rightarrow Df}(p,k,p_j) \approx\sum_{r^{\prime}}\mathcal{M}_{r^{\prime}}^{pp_t\rightarrow f}(p^{\prime},p_j) \,\Big(\frac{V_{r^{\prime}r}^D}{2k\cdot p-m_D^2}\Big).
\end{equation}
Here we have defined the vertex functions $V_{r^{\prime}r}^{S} =g_S\bar{u}^{r^{\prime}}(p^{{\prime}})u^r(p)$ and $V_{r^{\prime}r,\lambda}^{V} =g_V\bar{u}^{r^{\prime}}(p^{{\prime}})\cancel{\epsilon}^{\star}_{\lambda}(k)u^r(p)$, corresponding to dark scalar and dark vector radiation, respectively. Note that now the matrix element $\mathcal{M}^{pp_t\rightarrow f}$ involves the modified on-shell momentum $p^{\prime}\sim p-k$. 
Now by integrating over the phase space of the remaining particles in the final state ${X}$, the cross section for the process with dark state emission can be factorized as follows (see Appendix~\ref{app:ISR}),
\begin{align}\label{diffSplit}
    d\sigma^{pp_t\rightarrow Df}(s) \approx d\mathcal{P}_{p \rightarrow p^{\prime}D} \times\sigma_{pp}^{\rm NSD}(s^{\prime}),
\end{align}
where we have again introduced the differential splitting probability,
\begin{equation}
d\mathcal{P}_{p \rightarrow p^{\prime}D} \equiv w^{D}(z,p_T^2) d p_T^2 dz,
\end{equation}
for radiating a dark state given a longitudinal momentum fraction $z$ of the proton beam momentum and transverse momentum $p_T$. We take into account only the non-single diffractive cross section $\sigma_{pp}^{\rm NSD}(s^{\prime})$, since as discussed above radiation in quasi-elastic processes is suppressed by ISR and FSR interference. The residual scattering cross section involves $s^{\prime}=2m_p(p(1-z)+m_p)$ which accounts for the momentum of the emitted dark state. Deferring the details to Appendix~\ref{app:ISR}, the resulting splitting functions read
\begin{align}\label{SplitScalarMain}
 w_{S}(z,p_T^2)&=\frac{\alpha_\theta}{2\pi}F_S^2(m_S^2,m_p^2-H/z) \\
  & \times \frac{1}{2H} \bigg[z + z(1{-}z)\bigg(\frac{4m_p^2{-}m_S^2}{H}\bigg) \bigg], \nonumber
\end{align}
and 
\begin{align}\label{SplitVectorMain}
w_{V}(z,p_T^2) &= \frac{\alpha_{\epsilon}}{2\pi}F_V^2(m_V^2,m_p^2-H/z) \\
  &\times \frac{1}{H}\bigg[z-z(1{-}z) \Big(\frac{2m_p^2{+}m_{V}^2}{H}\Big) + \frac{H}{2z m_{V}^2}
\bigg] , \nonumber
\end{align}
with $\alpha_{\theta}=g_{SNN}^2\theta^2/4\pi$,  $\alpha_{\epsilon}=\alpha_{\rm em}\epsilon^2$, and the kinematic structure function is given by
\begin{equation}
H(z,p_T^2)\equiv p_T^2+z^2m_p^2+(1-z)m_{D}^2.
\end{equation}

The functions $F_S(k^2)$ and $F_V(k^2)$ are scalar and vector nucleon form factors, which are discussed in detail below in Section~\ref{subsec:FormFactor}. The expression in Eq.~(\ref{SplitScalarMain}) agrees with the result of Ref.~\cite{Boiarska:2019jym} in the case of scalar bremsstrahlung. 

Note that the terms appearing in Eqs.~(\ref{SplitScalarMain}) and (\ref{SplitVectorMain}) have poles when the the structure function $H(z,p_T^2)\rightarrow 0$. In the massless and collinear limit, the approximation gives the standard soft photon bremsstrahlung result with a $1/H\sim 1/p_T^2$ singularity \cite{Bjorken:2009mm}. Indeed, assuming a smooth limit as $m_V \rightarrow 0$, one restores the Altarelli-Parisi splitting kernel, $w_V \propto \frac{\alpha}{2\pi}\frac{1}{p_T^2}\frac{1{+}(1{-}z)^2}{z}$, for $m_p\ll p_T$. However, finite $m_{D}=m_V, m_S$ regulates this singularity, leading to $1/H \approx 1/m_D^2$. The final term in Eq.~(\ref{SplitVectorMain}) proportional to $1/m_V^2$ arises due to the longitudinal polarization of the massive vector, so the $m_V \rightarrow 0$ limit is not smooth, but this term is not numerically important for the parameters considered here.

The applicability of this approximation in which the ISR process is factorized from the underlying hard scattering depends on a number of kinematic conditions. 
First, we observe that the shift from the mass shell $p^{\prime 2}-m_p^2= (p{-}k)^{2}{-}m_p^2=0$ for the intermediate proton in the $pp^\star D$ vertex must be considerably smaller than the momentum transfer in the hard scattering denoted by the shaded central block of Fig.~(\ref{fig:ISR}). Fortunately, the required suppression when the intermediate proton line goes far off-shell  at the $pp^{\star}D$ vertex is accounted for by the off-shell (or transition) form-factor (\ref{FFoffshell}) described in the next section. The form factor is constructed to reach its maximum value ($=1$) if the invariant $p^{\prime 2}-m_p^2\approx -H/z$ is much less than a specified cut-off value $\Lambda_p \sim m_p$ associated with the hard scattering.  We will vary the hard scale over the range $1\lesssim \Lambda_p\lesssim 2~\gev$, with the central value of $1.5~\gev$, to assess the impact of this kinematic constraint.
As a second kinematic condition, the propagator approximation in Eq.~(\ref{PropApprox}) requires $E_{p^{\prime}}{-}(E_p{-}E_k)\ll 2E_{p^{\prime}}$. Along with the generic requirements that the process is relativistic with the beam energy being the dominant kinematic variable, this leads to two further kinematic consistency conditions,
\begin{align}\label{KinConditions}
&\frac{H}{4z(1-z)^2p_p^2} \ll 1,  \\
& p_T,\, m_p \, (m_D) \ll E_p \, (E_k). \label{kin2}
\end{align}  
For concreteness, we demand that the variable on the left of each inequality in Eqs. (\ref{KinConditions}) and (\ref{kin2}) be at most $20\%$ of the right hand side.

In combination, these kinematic conditions lead to a restricted range for $z$, as well as an upper bound on $p_T$ which depends on  $m_D$ and the characteristic scale $\Lambda_p$. As discussed further below, varying $\Lambda_p$ as described above leads to the red shaded bands in Figs.~\ref{fig:Production120GeV} and \ref{fig:Production14TeV}. This serves as an estimate for the theoretical uncertainty in our calculation, as the systematic uncertainty in the non-diffractive cross section in Eq.~(\ref{diffSplit}) is somewhat smaller.

\subsection{Time-like and off-shell form factors}\label{subsec:FormFactor}
The coherent emission of a dark vector or scalar from a proton, having timelike momentum, requires incorporation of a timelike form-factor to properly account for both the loss of coherence for momentum transfers above a GeV, and resonant enhancement due to mixing of the radiated dark state with hadronic degrees of freedom with the same quantum numbers. 

The vector case coincides with the proton electromagnetic form factor $F_{1,p}(q^2)$, which can be extracted from elastic scattering and annihilation reactions (see Ref.~\cite{Denig:2012by} for a recent review). Numerous data sets in the spacelike kinematic region have allowed high-precision parametrizations, but the timelike region is more complex and statistics over the kinematic threshold, $q^2 > 4m_p^2$, are limited~\cite{Adamuscin:2007iv,Fonvieille:2009px}, but it is this region involving low mass vector resonances that is of most interest to us. To make use of the data that exists, parametrizations in the low invariant mass regime have made use of the vector meson dominance (VMD) approach \cite{Adamuscin:2016rer,Faessler:2009tn} (see Ref.~\cite{TomasiGustafsson:2005kc} for a recent review). 
Following~\cite{dNCPR}, we make use of the following form-factor parametrization with a minimal number of free parameters which still achieves a good fit to data away from the threshold~\cite{Faessler:2009tn},
\begin{equation}\label{TLFFVector}
F_{1,V}^{p}(k_V^2)=\sum_{\rho,\omega}\frac{f_{\rho,\omega}m_{\rho,\omega}^2}{m_{\rho,\omega}^2-k_V^2-im_{\rho,\omega}\Gamma_{\rho,\omega}}.
\end{equation}
The fit parameters are $f_{\rho}=\{0.616,0.223,-0.339\}$, and $f_{\omega}=\{1.011,-0.881,0.369\}$, which account for mixing with $\rho$ and $\omega$ resonances.

Following Ref.~\cite{Batell:2020vqn}, lacking any data in the scalar channel, we take the same approach for the timelike scalar-nucleon form factor, incorporating mixing with isoscalar (and in principle isovector) $J^{PC} = 0^{++}$ scalar resonances through a sum of Breit-Wigner components,
\begin{equation}\label{TLFFScalar}
F_{1,S}^p(k_S^2)=\sum_{\phi=f_0}\frac{f_{\phi}m_{\phi}^2}{m_{\phi}^2-k_S^2-im_{\phi}\Gamma_{\phi}},
\end{equation}
where the parameters $f_{f_0}=\{0.28,1.8,-0.99\}$ account for mixing with the three low-lying scalar $f_0$ resonances.

The timelike form-factors assume all legs are on-shell. However, the intermediate proton in ISR and FSR is off-shell, and to account for the off-shell leg at the $pp^{\star}D$ vertex, as in Ref.~\cite{Feuster:1998cj} we introduce a further hadronic form factor,
\begin{equation}\label{FFoffshell}
F_{pp^{\star}D}(p^{\prime 2})=\frac{\Lambda_{p}^4}{\Lambda_{p}^4+(p^{\prime 2}-m_p^2)^2},
\end{equation}
which depends on the momentum of the intermediate proton $p^{\prime 2}=(p-k)^2$ rather than just the momentum transfer. The form factor is constructed to reach its maximum value ($=1$) if the invariant $p^{\prime 2}-m_p^2\approx -H/z$, which measures how far the intermediate proton line is off-shell, is much less than a specified cut-off value $\Lambda_p \sim m_p$. This off-shell hadronic form factor has been utilized in \textit{e.g.} $\pi$ and $\eta$ \cite{Ronchen:2012eg,Kamano:2013iva}, kaon \cite{Janssen:2001wk}, and $\omega$ \cite{Oh:2000zi} photoproduction reactions, and also in meson- and photon-induced reactions on the nucleon \cite{Penner:2002ma,Penner:2002md} to insure the gauge invariance of different contributions~\cite{Haberzettl:1998aqi}. We vary the hard scale over the range $1\lesssim \Lambda_p\lesssim 2~\gev$, with the central value of $1.5~\gev$, to generate the results shown in Figs.~\ref{fig:Production120GeV} and \ref{fig:Production14TeV}.
 
To account for both effects discussed above, we define the product of the timelike form-factors in Eqs.~(\ref{TLFFVector},\ref{TLFFScalar}) and the off-shell form factor in Eq.~(\ref{FFoffshell}), 
\begin{equation}\label{FormFactor}
F_D(k^2,p^{\prime 2}) \equiv F_{pp^{\star}D}(p^{\prime 2})\times F_{1,D}^p(k^2).
\end{equation}

\section{Results and comparisons}
In this section we present our results for the production rates of dark states via proton bremsstrahlung, and compare them with modifications of the Weizsacker-Williams approximation \cite{Kim:1973he,Bjorken:2009mm} which is particularly successful in modeling high energy electron bremsstrahlung.

\begin{figure*}[t]
\centering
    \includegraphics[width=0.49\textwidth]{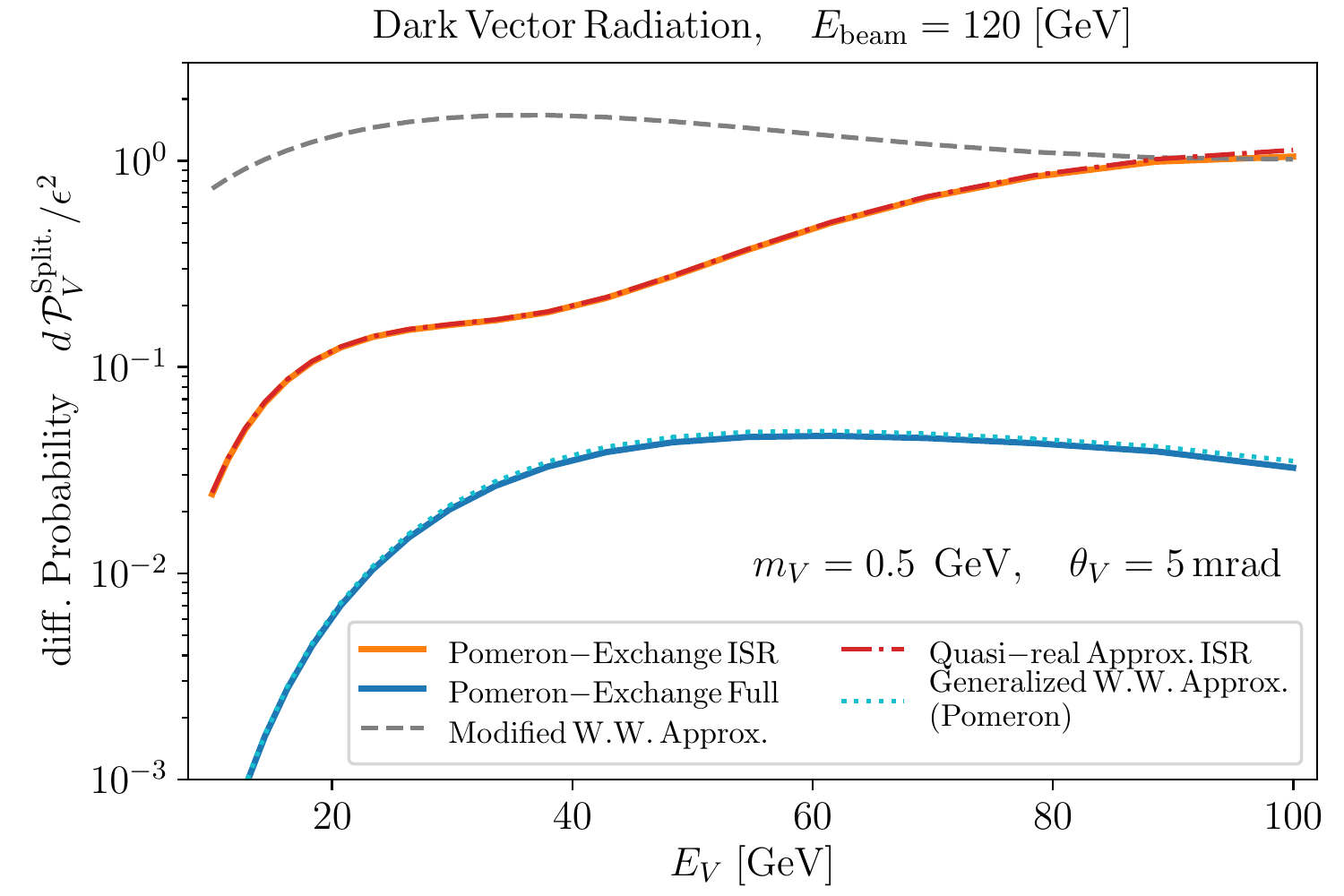}
    \includegraphics[width=0.49\textwidth]{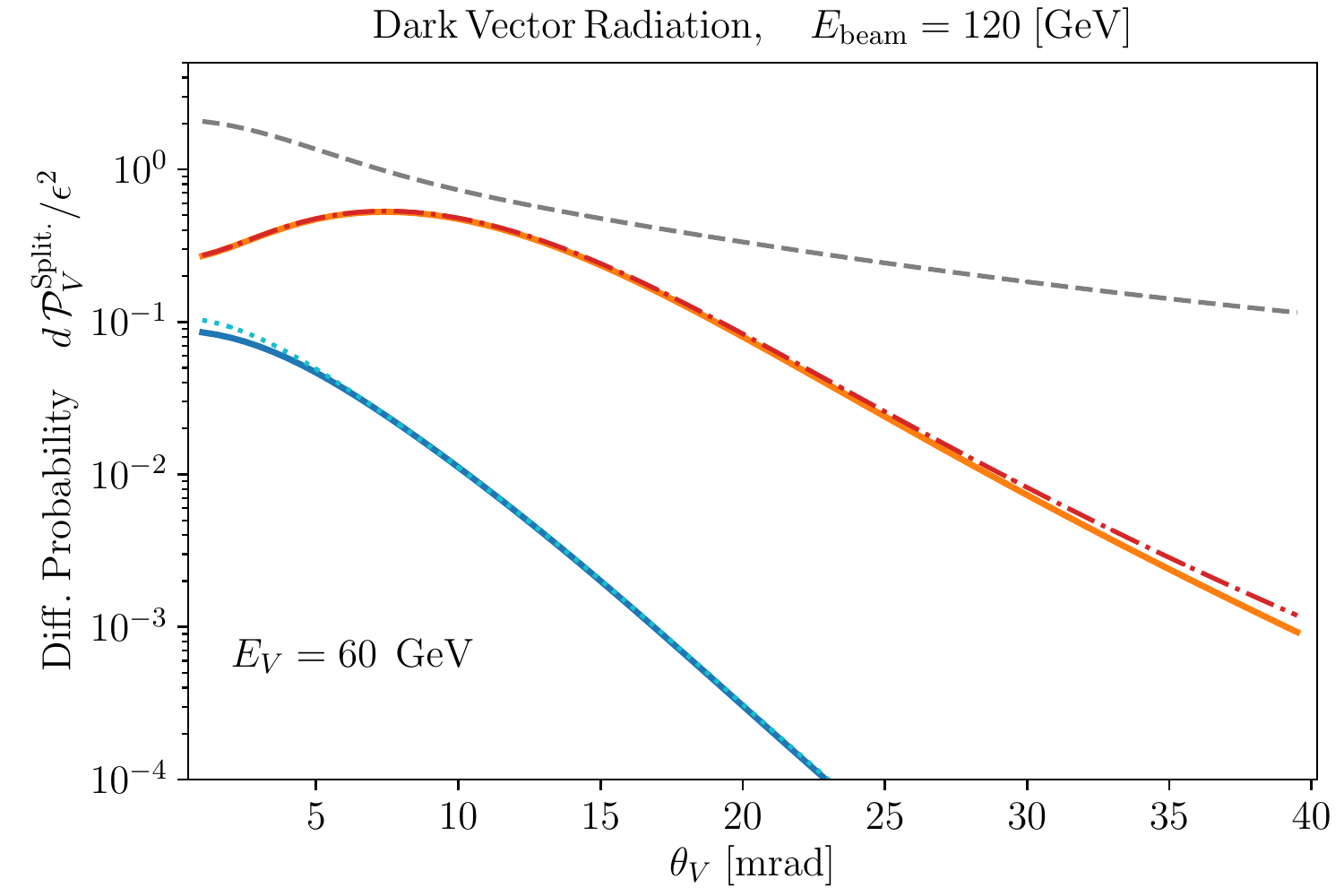}
    \includegraphics[width=0.49\textwidth]{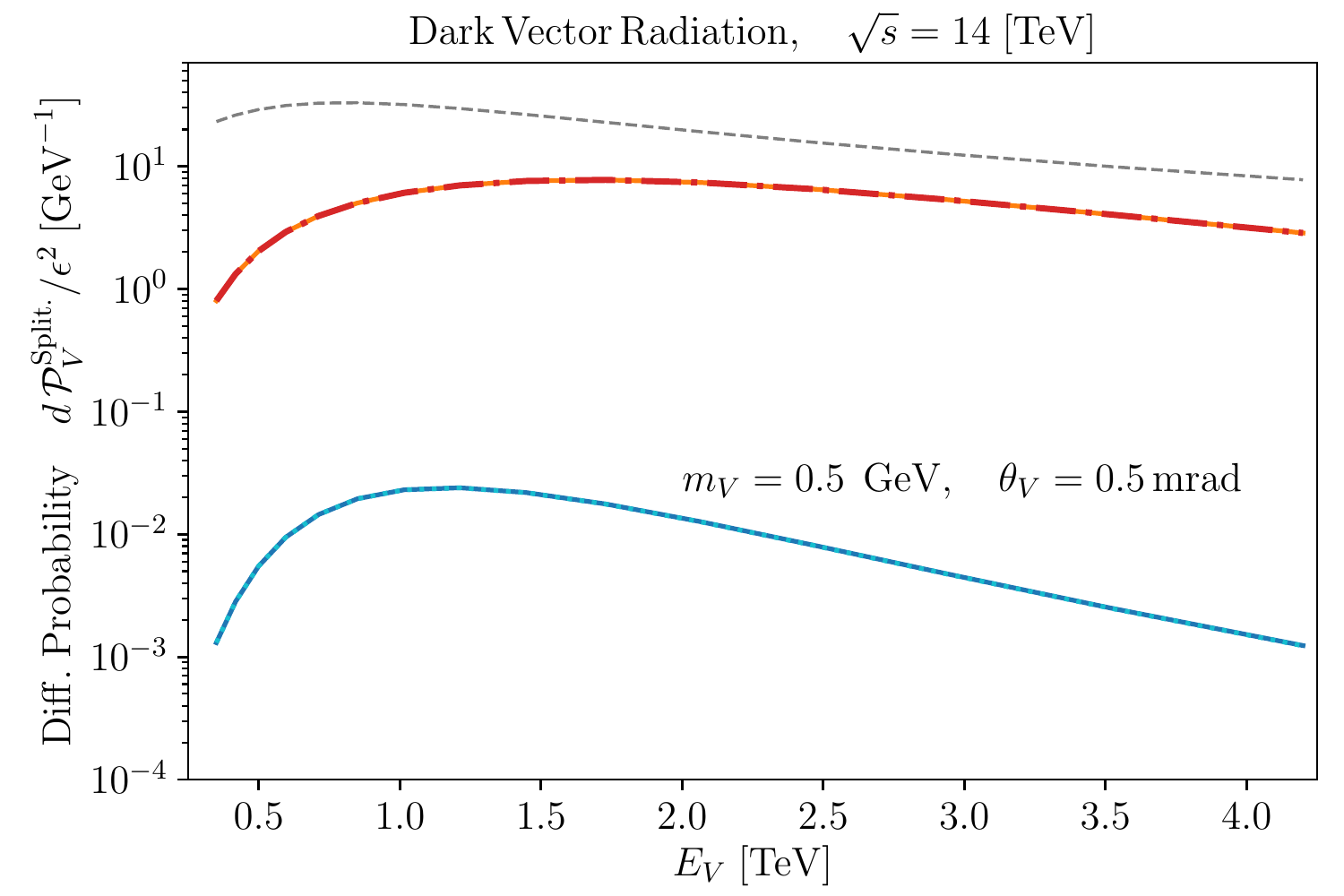}
    \includegraphics[width=0.49\textwidth]{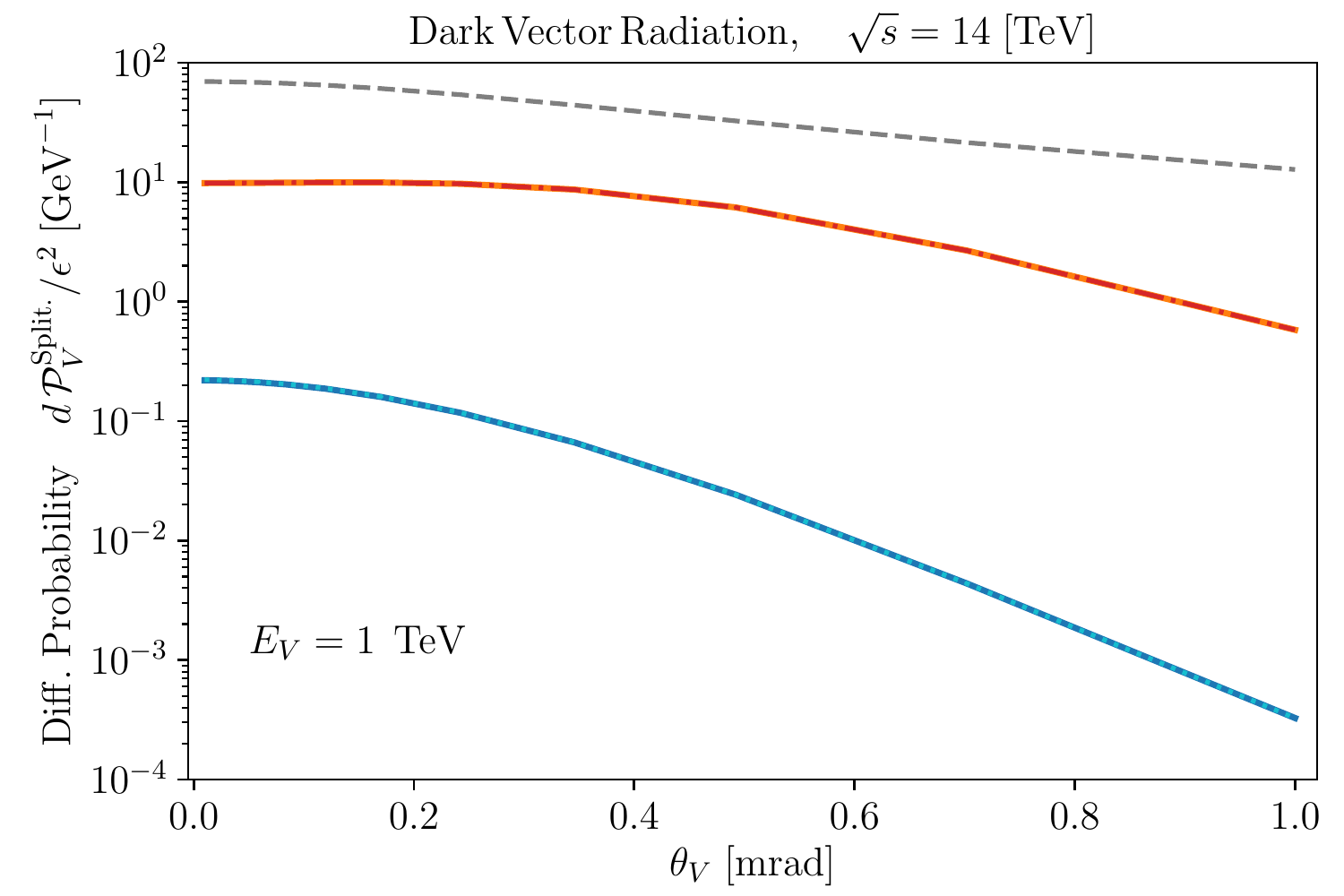}    
    \caption{The splitting probability for the proton to emit a dark vector with $m_V=0.5~\gev$ as a function of energy (left) and radiated angle (right), corresponding to beam energies $E_{\rm beam}=120~\gev$ and $\sqrt{s}=14~\tev$. The curves denote the contributions from the quasi-elastic emission from both initial and final state protons (solid blue), emission from initial state only (solid orange), compared with the approximate splitting probability using the WW approximation with an effective pomeron cloud (dotted cyan) and the quasi-real ISR (dashed-dotted red) methods. The latter curves overlap with the solid curves. Emission from both the initial and final state proton is subject to large interference and cancellations, in comparison to purely initial state radiation. The modified WW approximation of \cite{Blumlein:2013cua} is also shown for comparison (dashed gray).
}
    
\vspace{-0.5cm}
\label{fig:diffsplitVector}
\end{figure*}

\begin{figure*}[t]
\centering
    \includegraphics[width=0.49\textwidth]{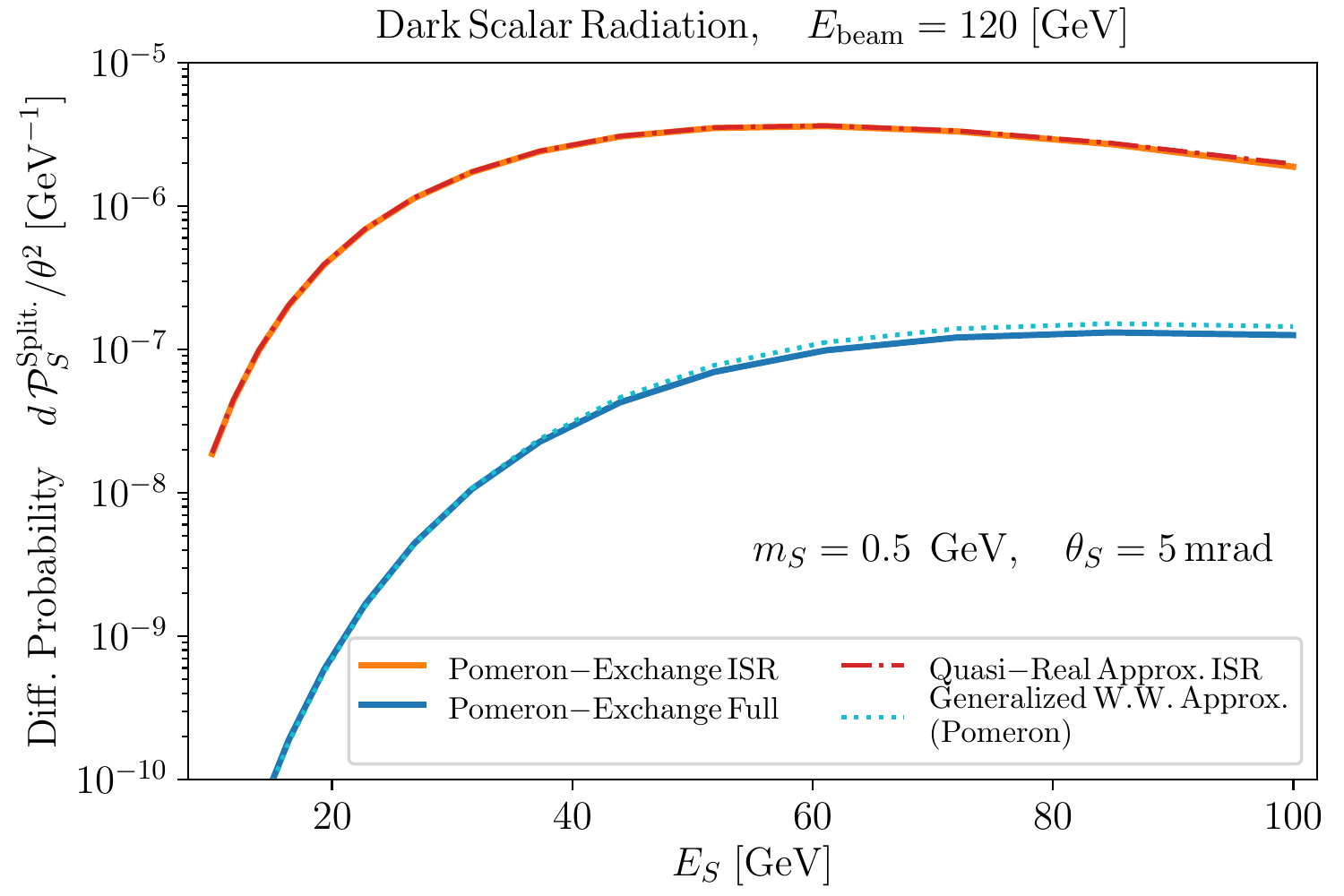}
    \includegraphics[width=0.49\textwidth]{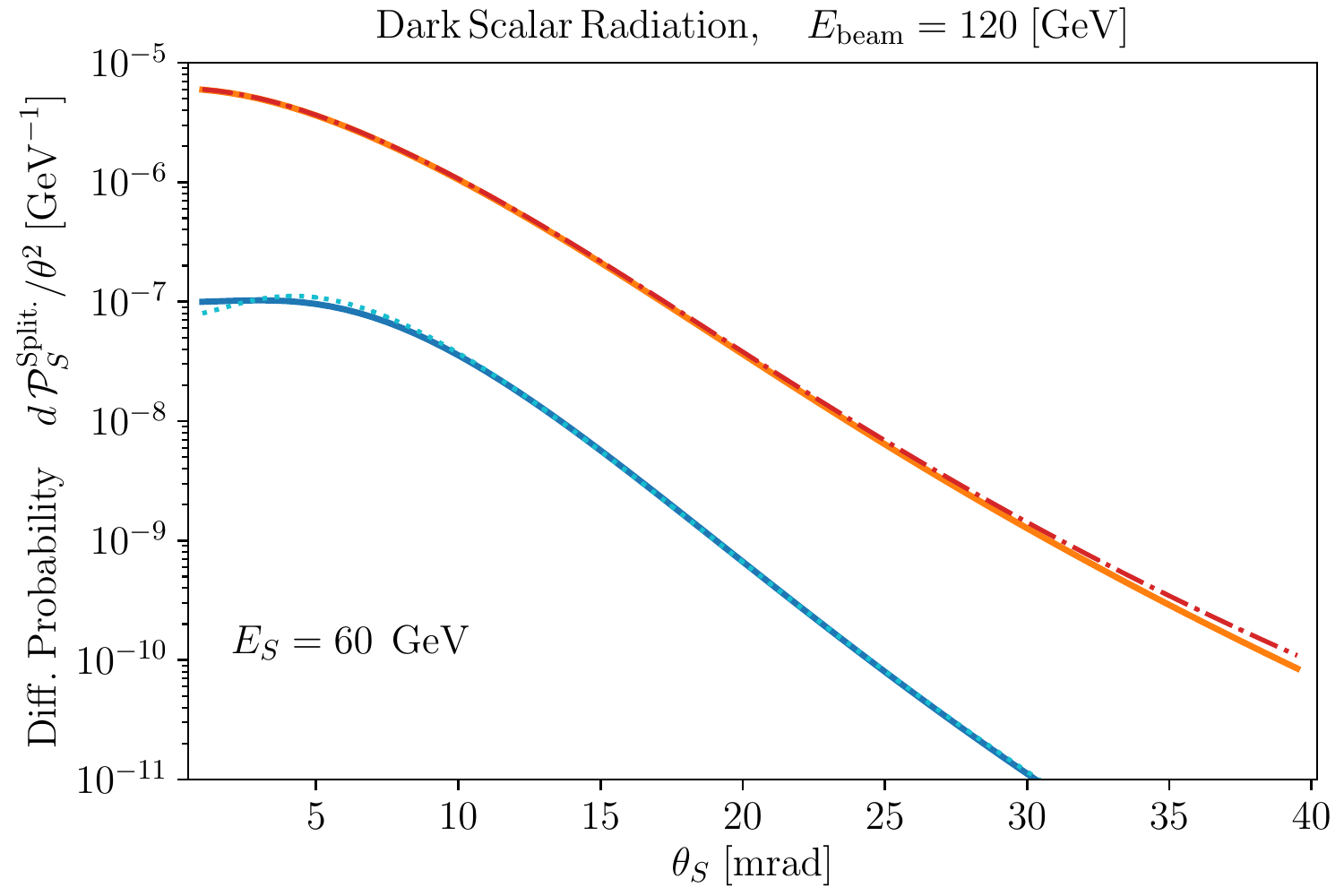}
        \includegraphics[width=0.49\textwidth]{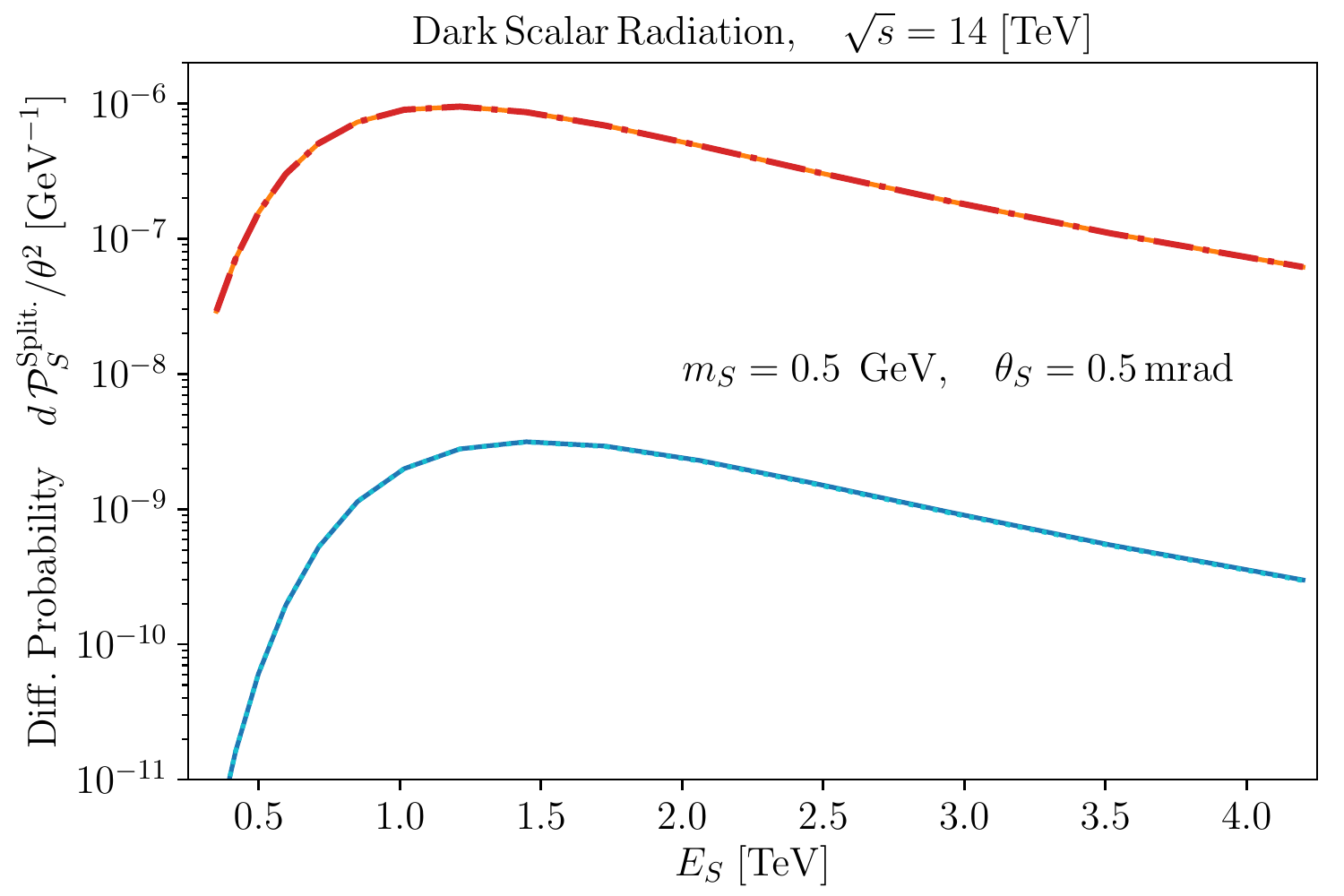}
    \includegraphics[width=0.49\textwidth]{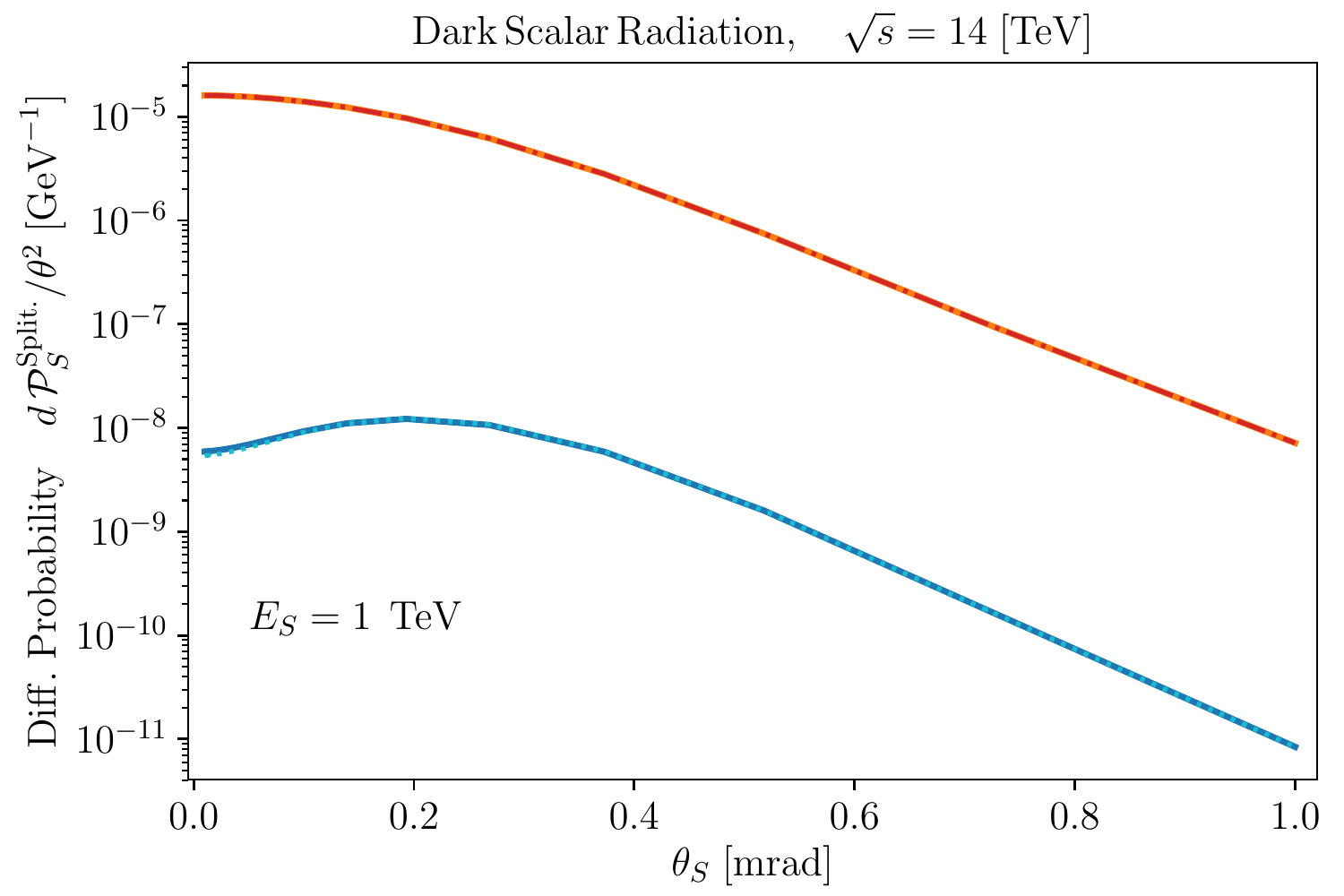}
    \caption{The splitting probability of the proton to emit a dark scalar with $m_S=0.5~\gev$ as a function of energy (left) and radiated angle (right), corresponding to beam energies $E_{\rm beam}=120~\gev$ and  $\sqrt{s}=14~\tev$. The curves denote the contributions from the quasi-elastic emission from both initial and final state protons (solid blue), emission from initial state only (solid orange), compared with the approximate splitting probability using the WW approximation with an effective pomeron cloud (dotted cyan) and the quasi-real ISR (dashed-dotted red) methods. The latter curves again overlap with the solid curves}
\vspace{-0.5cm}
\label{fig:diffsplitScalar}
\end{figure*}

Using the results of the last section, differential splitting probabilities for the various approaches are shown in Figs.~\ref{fig:diffsplitVector} and \ref{fig:diffsplitScalar} for vector and scalar dark sector radiation respectively. These results for the differential splitting functions illustrate a number of features. In particular, we see that the quasi-real approximation for ISR agrees very well with the full 2 to 3 calculation of pure ISR in the pomeron exchange model. We emphasize that although these processes in principle involve distinct final states, the ratio taken in forming the splitting functions restores the normalization. Similarly, we see that the full 2 to 3 calculation of ISR plus FSR in quasi-elastic scattering using the pomeron exchange model is well described by a hadronic generalization of the WW approximation described below. Finally, we have included the result from another modification of the WW approximation, presented in \cite{Blumlein:2013cua}, which we also discuss below and which leads to a slightly higher production rate. The form factor at the radiation vertex, as defined in Eq.~(\ref{FormFactor}), is used in all approaches except for the modified WW approximation, where following \cite{Feng:2017uoz} we use only a time-like form factor but restrict the transverse momentum of the radiated dark vector, $p_T < 1~\gev$.

Integrating these differential distributions, and incorporating the appropriate form-factors, our final results for the production rates of dark vectors and scalars are shown in Figs.~\ref{fig:Production120GeV} and \ref{fig:Production14TeV}. For illustrative purposes, in these figures, we choose angular cuts that are relevant for dark sector production in a 120 GeV fixed target beam at the Fermilab SeaQuest detector $\sim 10 $ m away from the beam collision, and at FASER in the forward region of the 14 TeV LHC.

\subsection{Versions of the WW approximation}\label{sec:WW}
The equivalent photon method (or Fermi-Weizsacker-Williams (WW) approximation) has been used successfully as an approximate method to evaluate cross sections for various QED processes at high energies (see e.g. \cite{Kim:1973he,Baier:1980kx,BERESTETSKII1982354}), wherein one replaces the target charge with its effective electromagnetic field. In this method, which is based on the mediator pole approximation, the 2 to 3 cross section is dominated by the photon pole at small $t$ corresponding to small photon virtuality. This approach has successfully been used for electron beams~\cite{Bjorken:2009mm,Berlin:2018bsc,Liu:2016mqv}, where in this case the scattering of the highly energetic beam electrons off the target reduces to a real photon interaction with the same target. 

Generalizing this argument to high energy proton-proton elastic scattering, we would have a cloud of effective bosons (here hypothetical pomerons) from the target proton denoted by $\chi_{\mathbb{P}}$, which the beam proton scatters to radiate a collinear dark state. In this case the cross-section for the full process can be expressed in terms of the 2 to 2 process $p{+}\mathbb{P}\rightarrow p^{\prime}{+}D$, a subprocess of the full 2 to 3 interaction. Following \cite{Kim:1973he,Bjorken:2009mm,Liu:2016mqv} we have,
\begin{align}\label{Brem_Elastic_CrossSection_WW}
    \Bigg( \frac{d\sigma^{\rm el}_{pp\rightarrow ppD}}{dzdp_T^2} \Bigg)_{\rm WW} & \cong \frac{\alpha_D}{16\pi^2}F_D^2(m_D^2,m_p^2-H/z)\frac{z(1-z)}{H^2} \nnl
    & \qquad \qquad \times(A^{22}_{D}|_{t=t_{min}})\, \chi_{\mathbb{P}},
\end{align}
with 
\begin{equation}
\chi_{\mathbb{P}}\equiv \int_{t_{min}}^{t_{max}}dt \,(t-t_{min})|\mathcal{A}_{el}(s,t)|^2,
\end{equation}
where $t_{min}{\approx} {-}H^2/(2z(1{-}z)p_p)^2$, $t_{max}{\approx }{-}2(1{-}z)m_pp_p$, and we have replaced $1/t$ for the photon propagator with the effective pomeron exchange amplitude, $\mathcal{A}_{\rm el}(s,t)$ defined in Eq.~(\ref{PomeronAmp}). Finally,  $A^{22}_{D}$ at $t{=}t_{min}$ corresponds to the 2 to 2 process and is presented in Appendix D. In Figs.~\ref{fig:diffsplitVector} and \ref{fig:diffsplitScalar} we show the splitting probability using this generalized WW approximation compared to the approaches described above. The Jacobian $\partial (z,p_T^2)/\partial (E_{k},\cos \theta_{k})=2kE_k/p_p$ is taken into account for comparing the splitting probability in Eq.~(\ref{Split_Prob}) with the corresponding one obtained from Eq.~(\ref{Brem_Elastic_CrossSection_WW}). We observe that this approximation agrees very well with the full 2 to 3 calculation in quasi-elastic scattering using pomeron exchange, and indeed is suppressed by a similar interference of ISR and FSR contributions.

Next, we discuss a different variant of the WW approximation. The procedure outlined in Ref.~\cite{Blumlein:2013cua} is a \textit{modified} version of both fermion-pole~\cite{Baier:1973ms,Altarelli:1977zs} and photon-pole approaches~\cite{Kim:1973he} (used for electron beams), and applied to the process of dark vector radiation in a high-energy proton beam dump. In this prescription, the following splitting function,
\begin{align}
w_{V}(z,p_T^2) &=  \frac{\alpha_{\epsilon}}{2\pi}|F_{1,V}^p(m_V^2)|^2\frac{1}{H}\bigg[\frac{1{+}(1{-}z)^2}{z}
\\
& {-}2z(1{-}z) \big(\frac{2m_p^2{+}m_{V}^2}{H}{-}z^2\frac{2m_p^4}{H^2}\big) 
\nnl
&
{+}2z(1{-}z)(1{+}(1{-}z)^2)\frac{m_p^2m_V^2}{H^2} 
{+}2z(1{-}z)^2\frac{m_V^4}{H^2}
\bigg], \nonumber
\end{align}
was determined using the matrix element for the WW sub-process $p+b\rightarrow p^{\prime}+V$, where the nature of the exchanged vector boson $b$ was not specified in \cite{Blumlein:2013cua}, but the pomeron is a viable candidate. This result notably includes terms of up quartic order in the mass scales and reduces to the well-known Altarelli-Parisi function in the massless limit. The splitting function was then convoluted with the total $pp$ cross section $\sigma_{pp}^{\rm tot}(s^{\prime})$ at a reduced scale $s^{\prime} = 2m_p(p_p(1-z)+m_p)$, and this approach, augmented with a timelike form-factor to account for mixing with hadronic states, has been widely used in estimating the bremsstrahlung production of dark vectors in recent years \cite{dNCPR,Feng:2017uoz,Berlin:2018jbm,Berlin:2018pwi}. We present the splitting probability from this prescription in Fig.~\ref{fig:diffsplitVector} to compare with the other approaches discussed in this paper. We observe that this rate is slightly higher than the one obtained using the quasi-real approximation for ISR. We also impose a constraint on either $p_T<1~\gev$ or the maximum angle from the beam $\theta_V$ when presenting the corresponding results in Figs.~\ref{fig:Production120GeV} and \ref{fig:Production14TeV}.

\section{Concluding remarks}
\label{sec:outlook}

In this paper, we have revisited one of the primary production channels for dark sector mediators (kinetically mixed vectors and Higgs portal scalars) at proton beam facilities, namely proton bremsstrahlung. The production rate is nontrivial to estimate in the forward region as it involves nonperturbative QCD. However, it is also important for a number of fixed target and accelerator based searches for dark sectors and light dark matter (including at the proposed Forward Physics Facility at the HL-LHC), as this production channel is enhanced by resonant hadronic mixing in the 0.5 - 1.0 GeV mass range. The analysis in this paper has focused solely on the production rates for the dark mediators. However, the production distributions obtained can straightforwardly be convoluted with decay distributions for either visible or hidden final states in the decays of dark vectors or scalars. 

Our approach has been to compare ISR and FSR channels in an explicit model for the underlying diffractive $pp$ scattering process with various approximations that instead use parametrizations of the hard scattering event. One of our primary goals was to compare the efficacy of these different approaches, and attempt to quantify the level of precision and distinct kinematic constraints. Our results for production distributions are summarized in Figs.~\ref{fig:diffsplitVector} and \ref{fig:diffsplitScalar}, while the final integrated results for the overall cross sections are shown in Figs.~\ref{fig:Production120GeV} and \ref{fig:Production14TeV}. The parameters chosen for these figures are representative of fixed target experiments close to the 120 GeV main injector beamline at Fermilab, and detectors such as FASER in the forward region of the ATLAS interaction point at the 14 TeV LHC. Overall, we find that radiation in quasi-elastic 2 to 3 scattering is suppressed by destructive interference between the $t$-channel ISR and FSR diagrams, while various approximations point to the dominant forward production channel being from ISR in non-single diffractive scattering.

\section*{Acknowledgements}

We would like to thank B. Batell, F. Kling, and R. Mammen-Abraham for helpful discussions and communication. This work was supported in part by NSERC, Canada.

\appendix

\renewcommand{\theequation}{\thesection\arabic{equation}}

\section{\texorpdfstring{$DL$}{Lg} Model for \texorpdfstring{$pp$}{Lg} scattering}\label{app:Pomeron}
In this appendix we review the DL parameterization for the $pp$ elastic scattering interaction at high energies, which includes Regge (exchange-degenerate $\rho, \omega$ and $f, a_2$ trajectories), single Pomeron exchange \cite{Donnachie:1984xq}, and multiple Regge and Pomeron exchanges.

Single-pomeron exchange describes the measured $pp$ total and elastic cross sections at high-energies, and small-$t$ processes reasonably well. We can present the model in terms of a phenomenological pomeron propagator $G_{\mathbb{P}}(s,t)g^{\mu\nu}$ and an effective proton-pomeron vertex $\Gamma_{\mathbb{P}}^{\mu}(t)$,
\begin{align} \label{propagator_single}
&G_{\mathbb{P}}(s,t)=\frac{(2\nu\alpha_{\mathbb{P}}^{\prime})^{\alpha_{\mathbb{P}}(t)}}{2\nu}\eta_{\mathbb{P}}(t),
\\
&\Gamma_{\mathbb{P}}^{\mu}(t)=-iY_{\mathbb{P}} F_{\mathbb{P}}(t) \gamma^{\mu},
\end{align}
where $2\nu = (s-u)/2$. The effective pomeron trajectory is linear in $t$,
\begin{equation}
\alpha_{\mathbb{P}}(t)=1+\epsilon_{\mathbb{P}}+\alpha_{\mathbb{P}}^{\prime}t,
\end{equation}
where the intercept $\alpha_{\mathbb{P}}(0)>1$, and $Y_{\mathbb{P}}$ is the coupling strength of the pomeron to the proton. The single pomeron form factor was traditionally assumed to have a dipole form~\cite{Donnachie:1983ff}, $F_{\mathbb{P}}(t)\sim 1/(1-t/0.71\, {\rm GeV}^2)^2$ analogous to the proton electromagnetic form factor. However, recent analyses~\cite{Donnachie:2013xia} have used an exponential form in the parametrization, $F_{\mathbb{P}}^2(t) = A\exp(at){+} (1{-}A)\exp(bt)$. Finally, $\eta_{\mathbb{P}}(t) {=}-\exp{(-\frac{1}{2}i\pi\alpha_{\mathbb{P}}(t))}$ is the signature factor. The relevant parameters are listed in Table~\ref{tab:fitPomeron}.

In principle one should also include the Reggeon $f_2$, $a_2$, $\rho$, and $\omega$ exchanges, but at large energies $\sqrt{s}\gg 10$ GeV these contributions with intercepts $\alpha_{\mathbb{R}}\simeq 0.5$ are suppressed, which allows us to focus on the Pomeron only \cite{Donnachie:1983ff}. Note that the electromagnetic contribution must also be included in order to fit elastic scattering data at very small $t$. However, this coulomb amplitude will be negligible for the parameters of interest here. 

\begin{table}[t]
\begin{center}
\begin{tabular}{ c||c } 
\hline
\hline
$\quad$ Parameter $\quad$   
& $\quad \quad$ Value $\quad \quad$
 \\
 \hline
 $\epsilon_{\mathbb{P}}$   & $0.110$
 \\ 
 $\alpha^{\prime}_{\mathbb{P}}$     & $0.165~\gev^{-2}$
 \\ 
 $Y_{\mathbb{P}}$          & $13.019$ 
 \\
 $Y_{3g}$          & $0.142$ 
 \\ 
 $t_0$          & $-4.230~\gev^2$
 \\ 
 $A$   & $0.682$
 \\ 
 $a$     & $7.854~\gev^{-2}$
 \\ 
 $b$          & $2.470~\gev^{-2}$
 \\ 
 \hline
 \hline
\end{tabular}
\end{center}
\caption{The best fit parameters for the elastic scattering DL model \cite{Donnachie:2013xia}.}
\label{tab:fitPomeron}
\end{table}

Measured high energy pp elastic scattering at the LHC \cite{Antchev:2011vs,Antchev:2011zz} may also point toward an additional hard-pomeron contribution \cite{Donnachie:2011aa}, although it is not strictly necessary to fit the data \cite{Donnachie:2013xia,Donnachie:2019ciz}. Instead, multiple-pomeron exchange, which one has to take into account to avoid the breakdown of unitarity, effectively behaves as a simple power of $s^{\epsilon}$ over a very wide range of energies. 
Following Ref.~\cite{Donnachie:2013xia}, we model the double-pomeron effective propagator and vertex as 
\begin{align}\label{propagator_Double}
&G_{\mathbb{PP}}(s,t) = \frac{(2\nu\alpha_{\mathbb{P}}^{\prime})^{\alpha_{\mathbb{PP}}(t)}}{2\nu}\eta_{\mathbb{PP}}(t),
\\
&\Gamma_{\mathbb{PP}}^{\mu}(t)=-iY_{\mathbb{PP}} F_{\mathbb{PP}}(t) \gamma^{\mu},
\end{align}
with the corresponding trajectory,
\begin{equation}
\alpha_{\mathbb{PP}}(t)=1+2\epsilon_{\mathbb{P}}+\frac{1}{2}\alpha_{\mathbb{P}}^{\prime}t,
\end{equation}
where the effective coupling strength of the double pomeron exchange takes the form $Y_{\mathbb{PP}}=Y_{\mathbb{P}}^2/4\sqrt{\pi}$. The signature factor reads $\eta_{\mathbb{PP}}(t) {=}\exp{(-\frac{1}{2}i\pi\alpha_{\mathbb{PP}}(t))}$, and the form factor squared takes the semi-eikonal form,
\begin{equation}
F_{\mathbb{PP}}^2(t) = \frac{A^2}{a/\alpha_{\mathbb{P}}^{\prime}+L}e^{\frac{1}{2}at}+\frac{(1-A)^2}{b/\alpha_{\mathbb{P}}^{\prime}+L}e^{\frac{1}{2}bt},
\end{equation}
with additional logarithmic factors $L =\log (2\nu\alpha_{\mathbb{P}}^{\prime})-i\pi/2$ in the denominator.

The dominant contribution to the elastic cross section at larger values of $|t| \gtrsim 3.5\, \mathrm{GeV}^2$ exhibits an energy-independent behavior $\sim t^{-8}$, in agreement with triple-gluon exchange \cite{Donnachie:1996rq}, with an amplitude of the from,
\begin{align}
&G_{3g}(t) = \begin{cases}
      \frac{|t_0|^3}{t^4}, & t<t_0 \\
      \frac{1}{|t_0|} e^{2(1-t^2/t_0^2)}, & t_0<t
    \end{cases} 
\\    
&\Gamma_{3g}^{\mu}=-iY_{3g}\gamma^{\mu},
\end{align}
where for $|t|<|t_0|$, a smooth transition was adopted to avoid a divergence as $t\rightarrow 0$.

The full parametrization involves adding these contributions, and the values of the best fit parameters of the DL model are given in the Table~\ref{tab:fitPomeron}.

\section{Matrix elements for quasi-elastic radiation}\label{app:MatEel}
In this appendix we present details of the matrix element calculation for the ISR plus FSR process,
\begin{equation} \label{emitprocess}
p(p_1) + p(p_2) \rightarrow p(p_3) + p(p_4) + D(k),
\end{equation}
where the dark state $D=\{S,V\}$ is emitted from either the beam or scattered proton with momentum $p_1$ and $p_3$, respectively (see e.g. \cite{Liu:2016mqv}). We define the Mandelstam variables $s{=}(p_1{+}p_2)^2$, $s^{\prime}{=}(p_1{+}p_2{-}k)^2$, and $t{=}q^2{=}(p_4{-}p_2)^2$. When $s \gg t$, the matrix elements for the bremsstrahlung processes depicted in Fig.~\ref{fig:brem_el} are given by,
\begin{align}
& i\mathcal{M}^{2\rightarrow 3}_S = i\bar{u}(p_4)\Gamma_{\mathbb{P}}^{\mu}(t)u(p_2) \bar{u}(p_3)S_{\mu}u(p_1), \\
& i\mathcal{M}^{2\rightarrow 3}_V = i\bar{u}(p_4)\Gamma^{\mu}_{\mathbb{P}}(t)u(p_2)\bar{u}(p_3)V_{\mu\alpha}u(p_1) \epsilon^{\alpha\star}_{k},
\end{align}
with 
\begin{widetext}
\begin{align}
S^{\mu} = ig_SF_{1,S}^p(m_S^2)\bigg[ & G_{\mathbb{P}}(s^{\prime},t)\Gamma^{\mu}_{\mathbb{P}}(t)F_{pp^{\star}D}\big((p_1{-}k)^2\big)\frac{i(\cancel{p}_{1}{-}\cancel{k}{+}m_p)}{(p_1{-}k)^2{-}m_p^2}+\frac{i(\cancel{p}_{3}{+}\cancel{k}{+}m_p)}{(p_3{+}k)^2{-}m_p^2}F_{pp^{\star}D}\big((p_3{+}k)^2\big)\Gamma^{\mu}_{\mathbb{P}}(t) G_{\mathbb{P}}(s,t)\bigg],
\end{align}
and
\begin{align}
V^{\mu\alpha} & = ig_VF_{1,V}^p(m_V^2)\bigg[ G_{\mathbb{P}}(s^{\prime},t)\Gamma^{\mu}_{\mathbb{P}}(t)F_{pp^{\star}D}\big((p_1{-}k)^2\big)\frac{i(\cancel{p}_{1}{-}\cancel{k}{+}m_p)}{(p_1{-}k)^2{-}m_p^2}\gamma^{\alpha} +\gamma^{\alpha}\frac{i(\cancel{p}_{3}{+}\cancel{k}{+}m_p)}{(p_3+k)^2{-}m_p^2}F_{pp^{\star}D}\big((p_3{+}k)^2\big)\Gamma^{\mu}_{\mathbb{P}}(t)G_{\mathbb{P}}(s,t)\bigg],
\end{align}
\end{widetext}
where the effective pomeron propagator $G_{\mathbb{P}}$ and vertex function $\Gamma^{\mu}_{\mathbb{P}}$ were introduced in Eq.~(\ref{propagator_single}), $g_V$ and $g_S$ are the effective dark vector and scalar couplings to the proton, and $\epsilon^{\mu}$ is the final state dark vector polarization with $\sum_{\rm pol}\epsilon^{\mu}_k \epsilon^{\nu}_k{=}{-}g^{\mu\nu}{+}k^{\mu}k^{\nu}/m_V^2$. When considering radiation from the proton $p$, the time-like vector and scalar proton form factors, $F_{1,V}^p(m_V^2)$ and $F_{1,S}^p(m_S^2)$, as well as the off-shell form factor $F_{pp^{\star}D}$, can be introduced in the form presented in~Section~\ref{subsec:FormFactor}.

The matrix element, averaged over spins and summed over the initial spins takes the form,
\begin{equation}
|\overline{\mathcal{M}_D^{2\rightarrow 3}}|^2 = \frac{1}{4}\sum_{\rm spin}|\mathcal{M}_D^{2\rightarrow 3}|^2 = T_{\mu\nu}B^{\mu\nu}_D \, ,
\end{equation}
where the initial spin-averaged target and beam proton tensors are given by,
\begin{align}
& T^{\mu\nu} = \frac{1}{2}\mathrm{Tr}\big[ (\cancel{p}_4+m_p)\Gamma^{\mu}_{\mathbb{P}}(\cancel{p}_2+m_p)\Gamma^{\nu\star}_{\mathbb{P}}\big], \\
& B^{\mu\nu}_V = \frac{1}{2}\mathrm{Tr}\big[ (\cancel{p}_3+m_p)V^{\mu\alpha}(\cancel{p}_1+m_p)V^{\beta\nu\star}\big] \nonumber\\
 & \qquad\quad\qquad \times (-g_{\alpha\beta}+\frac{k_{\alpha}k_{\beta}}{m_V^2}), \\
& B^{\mu\nu}_S = \frac{1}{2}\mathrm{Tr}\big[ (\cancel{p}_3+m_p)S^{\mu}(\cancel{p}_1+m_p)S^{\nu\star}\big].
\end{align} 

The differential cross section for the process (\ref{emitprocess}) in the lab-frame,
\begin{align}
d\sigma &= \frac{1}{4|\vec{p}_1| m_p}|\overline{\mathcal{M}_D^{2\rightarrow 3}}|^2(2\pi)^4\delta^{(4)}(p_1+q-k-p_3)
\nonumber \\ 
& \times \frac{d^3\vec{p}_3}{(2\pi)^3 2E_3}\frac{d^3\vec{p}_4}{(2\pi)^3 2E_4}\frac{d^3\vec{k}}{(2\pi)^3 2E_k} \, ,
\end{align}
takes the following form after integrating over $\vec{p_3}$ and changing variables from
$\vec{p_4}$ to $\vec{q}$ (the three-momentum of $q^{\mu}=(p_2^{\mu}{-}p_4^{\mu})$), and using the remaining $\delta$-function to integrate out $|\vec{q}|$, 
\begin{equation}
d\sigma = \frac{|\mathcal{M}_D^{2\rightarrow 3}|^2 \, d^3\vec{k}}{32(2\pi)^5|\vec{p}_1| m_p E_k}\frac{Q^2 \, d\cos\theta_qd\phi_q}{\big[Q(E_3{+}E_4){-}RE_3\cos\theta_q\big]} \, .
\end{equation}
Here $\theta_q$ and $\phi_{q}$ are the polar and azimuthal angles of $\vec{q}$ in the direction of $\vec{R}=\vec{k}-\vec{p}_1$, $Q=|\vec{q}|$, $R=|\vec{R}|$, and $\cos\theta_q=(m_p^2{+}R^2{+}Q^2{-}E_3^2)/(2QR)$.

Next we change variables from $\theta_q$ to $t=q^2$ to obtain
\begin{align}\label{appElCrossSectionLab}
\frac{d^2 \sigma}{dE_k d\cos \theta_k } &= \frac{1}{64(2\pi)^4 p_1 m_p^2}\frac{|\vec{k}|}{|\vec{p}_1{-}\vec{k}|}
\nonumber \\ 
& \times \int_{t_{min}}^{t_{max}} dt \int_{0}^{2\pi} d\phi_q\, |\overline{\mathcal{M}^{pp\rightarrow ppD}}|^2 \, ,
\end{align}
where $t = 2m_p(m_p-\sqrt{Q^2+m_p^2})$, and the integration boundaries of $t$ correspond to 
\begin{equation}
[Q]^{\pm} = \frac{R}{2} \pm \frac{(E_1{-}E_k{+}m_p)}{2}\sqrt{1-\frac{4m_p^2}{(E_1{-}E_k{+}m_p)^2{-}R^2}}.
\end{equation}
The equivalent center of mass (CM) version of Eq.~(\ref{appElCrossSectionLab}) has a similar form,
\begin{align}\label{appElCrossSectionCM}
\frac{d^2 \sigma}{dE_k^{\star} d\cos \theta_k^{\star} } &= \frac{1}{64(2\pi)^4 p_{\rm CM} E_{\rm CM}^2}
\nonumber \\ 
& \times \int_{\tilde{t}_{min}}^{\tilde{t}_{max}} d\tilde{t} \int_{0}^{2\pi} d\phi_{\tilde{q}}\, |\overline{\mathcal{M}^{pp\rightarrow ppD}}|^2 \, ,
\end{align}
where we defined the invariant $\tilde{t}\equiv (p_1{+}p_2{-}p_4)^2=m_p^2+E_{\rm CM}(E_{\rm CM}-2\sqrt{\tilde{Q}^2+m_p^2})$, and starred $(^\star)$ variables are defined in the CM frame. The integration boundaries for $\tilde{t}$ correspond to
\begin{equation}
[\tilde{Q}]^{\pm} = \frac{k^{\star}}{2} \pm \frac{(E_{\rm CM}{-}E_k^{\star})}{2}\sqrt{1-\frac{4m_p^2}{(E_{\rm CM}{-}E_k)^2{-}k^2}}.
\end{equation}

\section{Initial State Radiation}\label{app:ISR}

Consider the process of scattering of a beam proton by a target proton $pp_t\rightarrow X$, where $X$ denotes any number of charged (and neutral) particles in the final state. Along with this process, let us consider another process, $pp_t \rightarrow f+D$, involving an additional dark state $D$ which is emitted from the incoming proton as shown in Fig.~\ref{fig:ISR}. Under certain kinematic conditions (formulated below), the cross-section for the second process can be represented as a product of two independent factors, the cross-section of the former sub-process and the splitting probability for the emission of a single dark state in the collision.

The corresponding amplitude for the hard process $pp_t\rightarrow X$ can be written as
\begin{equation}
    \mathcal{M}_r^{pp_t\rightarrow f}=A(p,p_j)u^r(p),
\end{equation}
where $u^r(p)$ is the spinor of the incoming proton with the helicity $r$ and momentum $p$, and $A(p,p_j)$ the remaining part of the amplitude for the hard scattering with $p_j$ denoting the momenta of the other particles in the scattering.

For the process of dark state emission  from the incoming proton in the initial state, the matrix element can be obtained from that for the original process by adding an external dark state line, and reads
\begin{align}\label{propagator_ISR}
&\mathcal{M}_r^{pp_t\rightarrow Sf}(p,k,p_j)  \nonumber\\
 & \qquad =ig_S A(p{-}k,p_j)\frac{i(\cancel{p}{-}\cancel{k}{+}m_p)}{(p{-}k)^2{-}m_p^2}u^r(p), \\
&\mathcal{M}_{r\lambda}^{pp_t\rightarrow Vf}(p,k,p_j) \nonumber \\
& \qquad=ig_V A(p{-}k,p_j)\frac{i(\cancel{p}{-}\cancel{k}{+}m_p)}{(p{-}k)^2{-}m_p^2}\cancel{\epsilon}^{\star}_{\lambda}(k)u^r(p), 
\end{align}
where $k$ and $p^{\prime}=p-k$ denote the four-momenta of the dark state and the internal
proton, respectively.

The kinematic variables in the infinite momentum frame can be parametrized as follows,
\begin{align}\label{momentum_ISR}
& p^{\mu} = (p_{p}+\frac{m_{p}^{2}}{2p_{p}},\mathbf{0},p_{p}), \nonumber\\
& k^{\mu} = (zp_{p}+\frac{p_{T}^{2}+m_{D}^2}{2zp_{p}},\mathbf{p}_{T},zp_{p}),\\ 
& p^{\prime \,\mu} = ((1-z)p_{p}+\frac{p_{T}^{2}+m_{p}^{2}}{2p_{p}(1-z)},-\mathbf{p}_{T},(1{-}z)p_{p}) , \nonumber
\end{align}
where $z$ is the fraction of longitudinal momentum carried by D, and the transverse momentum $\mathbf{p}_{T}=(p_T\cos \phi, p_T\sin \phi)$ is a measure of the non-collinearity of the radiated scalar which determines how far off shell the intermediate proton is, as given by $(p-k)^{2}= m_p^2 - H/z$, where we have defined the kinematic structure function 
\begin{equation}
H(z,p_T^2)\equiv p_T^2+z^2m_p^2+(1-z)m_{D}^2.
\end{equation}

We use the  representation for the intermediate proton propagator of Eq.~(\ref{propagator_ISR}) in the framework of the quasi-real approximation which appears suitable for evaluating the cross section in the high energy limit. The intermediate proton propagator corresponding to two possible time orderings can be decomposed as follows,
\begin{align}\label{propagator_approx}
&\frac{i(\cancel{p}-\cancel{k}+m_p)}{(p-k)^2-m_p^2} = \frac{i}{2E_{p^{\prime}}}\sum_{r^{\prime}}
\\
&\qquad\qquad \bigg[ \frac{u^{r^{\prime}}(p{-}k)\bar{u}^{r^{\prime}}(p{-}k)}{E_p{+}E_k{-}E_{p^{\prime}}}+\frac{v^{r^{\prime}}({-}p{-}k)\bar{v}^{r^{\prime}}({-}p{-}k)}{E_p{-}E_k{+}E_{p^{\prime}}} \bigg], \nonumber
\end{align}
where $E_p$, $E_{p^{\prime}}$, and $E_k$ are defined in Eqs. (\ref{momentum_ISR}) as the energy of the incoming proton, the intermediate proton and the radiated dark state.

In the collinear limit, $k\cdot p/p_p^2 \ll 1$, where the dark states are radiated almost parallel to the energetic beam
proton, the denominator of the first term on the right hand side of Eq.~(\ref{propagator_approx}) is small relative to the denominator of the second term, $E_p{+}E_k{-}E_{p^{\prime}}\ll E_p{-}E_k{+}E_{p^{\prime}}$, which implies,
\begin{equation}
\frac{H}{4z(1-z)^2p_p^2} \ll 1.
\end{equation}
Provided that the above kinematic condition is satisfied, one can retain the first term while neglecting the second. Thus the numerator of the proton propagator in Eq.~(\ref{propagator_ISR}) can be replaced with the polarization sum of an on-shell fermion, with the result
\begin{align}\label{mat_elem_ISR}
\mathcal{M}_r^{pp_t\rightarrow Sf} & \approx      -A(p^{\prime},p_j)\sum_{r^{\prime}}u^{r^{\prime}}(p^{\prime})\frac{\Big(g_S\bar{u}^{r^{\prime}}(p^{{\prime}})u^r(p)\Big) }{(p{-}k)^2{-}m_p^2} \nonumber 
   \\
& =\sum_{r^{\prime}}\mathcal{M}_{r^{\prime}}^{pp_t\rightarrow f}(p^{\prime},p_j) \,\Big(\frac{z}{H}\Big) \, V_{r^{\prime}r}^S,
   \\
\mathcal{M}_{r,\lambda}^{pp_t\rightarrow Vf} & \approx      -A(p^{\prime},p_j)\sum_{r^{\prime}}u^{r^{\prime}}(p^{\prime})\frac{\Big(g_V\bar{u}^{r^{\prime}}(p^{{\prime}})\cancel{\epsilon}^{\star}_{\lambda}(k) u^r(p)\Big)}{(p{-}k)^2{-}m_p^2} \nonumber 
   \\
& =\sum_{r^{\prime}}\mathcal{M}_{r^{\prime}}^{pp_t\rightarrow f}(p^{\prime},p_j) \,\Big(\frac{z}{H}\Big) \, V_{r^{\prime}r,\lambda}^V,
\end{align}
where we defined the vertex functions $V_{r^{\prime}r}^{S} =g_S\bar{u}^{r^{\prime}}(p^{{\prime}})u^r(p)$ and $V_{r^{\prime}r,\lambda}^{V} =g_V\bar{u}^{r^{\prime}}(p^{{\prime}})\cancel{\epsilon}^{\star}_{\lambda}(k)u^r(p)$. Note that now the matrix element $\mathcal{M}^{pp_t\rightarrow f}$ involves the reduced momentum $p'\sim (1-z)p$.

To calculate the vertex functions above, we use the Pauli representations of the right-handed and left-handed helicity states $u^{r}(p)$, for $r=\pm$, with momentum $\textbf{p}=|\vec{p}|\,(\sin\theta\cos\phi,\sin \theta\sin\phi,\cos \theta)$, normalized to $u^{\dagger}u = 2E$ particles per unit volume, which take the form,
\begin{align}
    & u^{(+)}(p)= \sqrt{E+m} \Big(c,s\, e^{i\phi},\frac{|\vec{p}|}{E+m}c,\frac{|\vec{p}|}{E+m}s\,e^{i\phi} \Big)^T, \nonumber \\ 
    & u^{(-)}(p)= \sqrt{E+m}\Big({-}s,c\,e^{i\phi},\frac{|\vec{p}|}{E+m}s,\frac{-|\vec{p}|}{E+m}c\,e^{i\phi} \Big)^T,
\end{align}
where $s\equiv\sin (\theta/2)$ and $c\equiv\cos (\theta/2)$. The expressions for the circular and longitudinal polarization vectors associated to the dark vector with momentum $\textbf{k}=|\vec{k}|\,(\sin\theta\cos\phi,\sin \theta\sin\phi,\cos \theta)$ also read,
\begin{align}
    & \epsilon^{\mu}_{\pm}(k)= \frac{e^{\pm i \phi}}{\sqrt{2} }\Big(0,\mp \cos \theta \cos \phi{+}i\sin \phi, \nonumber\\
     & \qquad \qquad\qquad \qquad\qquad \mp\cos \theta\sin \phi {-}i\cos \phi,\pm \sin \theta \Big),\nonumber \\ 
    & \epsilon^{\mu}_{ L}(k)= \frac{1}{m_V}\Big(|\vec{k}|, E_k \sin \theta \cos \phi, E_k \sin \theta\sin \phi ,E_k \cos \theta \Big).
\end{align}
Thus, by using the kinematic variables defined in Eq.~(\ref{momentum_ISR}, one finds the following explicit expressions for the spinors 
\begin{align}
    & u^{+}(p) {=} (\sqrt{E_p{+}m_p},0,\frac{p_p}{\sqrt{E_p{+}m_p}},0)^T ,
    \\
   & u^{-}(p) {=} (0,\sqrt{E_p{+}m_p},0,\frac{{-}p_p}{\sqrt{E_p{+}m_p}})^T
    \nonumber,
    \\
    & u^{+}(p^{\prime}) {=} \sqrt{E_{p^{\prime}}{+}m_p}\bigg(1,\frac{p_T\,e^{i(\phi{+}\pi)}}{2(1{-}z)p_p},\frac{(1{-}z)p_p}{E_{p^{\prime}}{+}m_p}, \nonumber\\
     & \qquad\qquad\qquad\qquad\qquad\qquad\qquad \frac{\frac{1}{2} p_T \, e^{i(\phi{+}\pi)}}{E_{p^{\prime}}{+}m_p} \bigg)^T
    \nonumber,
    \\
    & u^{-}(p^{\prime}) {=} \sqrt{E_{p^{\prime}}{+}m_p}\bigg(\frac{-p_T}{2(1{-}z)p_p},e^{i(\phi{+}\pi)},\frac{\frac{1}{2} p_T}{E_{p^{\prime}}{+}m_p},
    \nnl
    &\qquad\qquad\qquad\qquad\qquad\qquad\qquad\frac{-(1{-}z)p_p\,e^{i(\phi{+}\pi)}}{E_{p^{\prime}}{+}m_p} \bigg)^T \nonumber ,
\end{align}
and the polarization vectors,
\begin{align}
    & \epsilon^{\mu}_{\pm}(k)= \frac{1}{\sqrt{2}}\Big(0,1,\pm i, -\frac{p_T}{zp_p} e^{\pm i\phi}\Big), \\ 
    & \epsilon^{\mu}_{ L}(k)= \frac{1}{m_V}\Big(zp_p, p_T \cos \phi, p_T\sin \phi ,zp_p \Big). \label{Pol}
\end{align}
Straightforward algebra then yields the following vertex functions up to $\mathcal{O}(m_p^2,p_T^2)$,
\begin{align}
    &V_{r^{\prime}r}^S = \frac{g_S}{\sqrt{1-z}}e^{i(\frac{r{-}1}{2})\phi}\Bigg[r(2-z)m_p\delta_{r^{\prime}r} -p_T \delta_{r^{\prime},-r}\Bigg], 
    \\
  &V_{r^{\prime}r,\lambda=\pm}^V  = \frac{g_V\sqrt{2}}{z\sqrt{1{-}z}}e^{-i\lambda\phi}e^{i(\frac{r{-}1}{2})\phi} \nonumber\\
  & \qquad \times \Bigg[ p_T \big((1{-}z)\delta_{r,-\lambda}{-}\delta_{r\lambda}\big) \delta_{r^{\prime}r} - \lambda z^2m_p \delta_{r\lambda}\delta_{r^{\prime},-r}\Bigg] , \\   
  &V_{r^{\prime}r,\lambda=L}^V  = \frac{g_V}{z\sqrt{1{-}z}}\frac{re^{i(\frac{r{-}1}{2})\phi}}{m_V} \nonumber\\
   &\qquad \times \bigg(p_T^2{+}z^2m_p^2{-}(1{-}z)m_V^2\bigg) \delta_{r^{\prime}r},
\end{align}
We then obtain
\begin{align}\label{Vtx_S}
V_{r^{\prime},r}^S \big(V_{r^{\prime\prime},r}^S\big)^{\star} &= g_S^2 \bigg(\mathcal{I}_S\,\delta_{r^{\prime}r} \delta_{r^{\prime\prime},r^{\prime}} \nonumber\\
 & +\mathcal{J}_S\,r^{\prime}\,\big(\delta_{r^{\prime}r}{-}\delta_{r^{\prime},-r}\big)\delta_{r^{\prime\prime},-r^{\prime}}\bigg),
\end{align}
and
\begin{align}\label{Vtx_V}
 \sum_{\lambda=\pm,L} V_{r^{\prime},r,\lambda}^V \big(V_{r^{\prime\prime},r,\lambda}^V\big)^{\star} & =  g_V^2 \bigg( \mathcal{I}_V\,\delta_{r^{\prime}r} \delta_{r^{\prime\prime},r^{\prime}} \\
& +\mathcal{J}_V\,r^{\prime}\,\big(\delta_{r^{\prime}r}{-}\delta_{r^{\prime},-r}\big)\delta_{r^{\prime\prime},-r^{\prime}}\bigg), \nonumber
\end{align}
which involves the functions,
\begin{align}
\mathcal{I}_S & = \frac{(2{-}z)^2 m_p^2+p_T^2}{1{-}z}, \nnl
\mathcal{J}_S & =  -\frac{(2{-}z)}{1{-}z}m_p p_T, \nnl
\mathcal{I}_V & =  \frac{2}{(1{-}z)}\bigg(\frac{1{+}(1-z)^2}{z^2}p_T^2+z^2m_p^2 \nnl
& \quad\quad\quad +\frac{1}{2z^2m_V^2}\Big(p_T^2{+}z^2m_p^2{-}(1{-}z)m_V^2\Big)^2\bigg), \nnl
\mathcal{J}_V & =  0,
\end{align}
which are independent of $\phi$. Note that the last term in $\mathcal{I_{V}}$ arises only for the massive vector boson, due to the longitudinal polarization given in (\ref{Pol}).

Inserting the vertex functions (\ref{Vtx_S}) and (\ref{Vtx_V}) into Eq.~(\ref{mat_elem_ISR}), the absolute square of the matrix element summed over polarizations of the final proton can now be expressed in the form,
\begin{align} \label{MatElemSum}
    &\frac{1}{2}\sum_{r(,\lambda)}|\mathcal{M}_{r(,\lambda)}^{pp_t\rightarrow Df} (p,k,p_j)|^2  \nonumber\\
    & = \big(\frac{z}{H}\big)^2 \sum_{r,r^{\prime},r^{\prime \prime}(,\lambda)}V_{r^{\prime},r(,\lambda)}^D \big(V_{r^{\prime\prime},r(,\lambda)}^D)^{\star}\nnl
    & \quad \quad \quad \quad  \times\frac{1}{2} \mathcal{M}_{r^{\prime}}^{pp_t\rightarrow f}(p^{\prime},p_j)\bigg(\mathcal{M}_{r^{\prime\prime}}^{pp_t\rightarrow f}(p^{\prime},p_j)\bigg)^{*}  \nonumber\\
    & = g_D^2 \big(\frac{z}{H}\big)^2 \mathcal{I}_D|\overline{\mathcal{M}^{p^{\prime}p_t\rightarrow f}}|^2,
\end{align}
which importantly is proportional to the matrix element squared for the subprocess $pp_t \rightarrow X$. Note that terms with $\big(\delta_{r^{\prime}r}{-}\delta_{r^{\prime},-r}\big)\delta_{r^{\prime\prime},-r^{\prime}}$ in Eqs. (\ref{Vtx_S}) and (\ref{Vtx_V}) vanish. This is because collinear emission does not change the proton's helicity and only transitions in which the helicity is conserved contribute to the unpolarized matrix element in Eq.~(\ref{MatElemSum}).

Finally, by integrating over the final state phase space, the cross section for dark state emission is expressed via the unpolarized cross section of the sub-process $pp_t \rightarrow X$ without radiation at the reduced momentum $(1-z)p_p$,
\begin{widetext}
\begin{align}
    d\sigma^{pp_t\rightarrow Df}(s) &=\frac{1}{4E_p E_{p_t}}\frac{d^3k}{(2\pi)^32E_k} \prod_f \frac{d^3p_f}{(2\pi)^3 2E_f}|\overline{\mathcal{M}^{p^{\prime}p_t\rightarrow Df}}|^2 (2\pi)^4\delta(p{+}p_{t}{-}k{-}p_f) \nonumber \\
    &\approx g_D^2\big(\frac{z}{H}\big)^2 \mathcal{I}_D \frac{dp_T^2dz}{16\pi^2z} \frac{E_{p^{\prime}}}{E_p} \int \frac{1}{4E_{p^{\prime}} E_{p_t}}\prod_f \frac{d^3p_f}{(2\pi)^3 2E_f}|\overline{\mathcal{M}^{p^{\prime}p_t\rightarrow f}}|^2 (2\pi)^4\delta(p^{\prime}{+}p_{t}{-} p_f) \nonumber \\
    & \approx w_{D}(z,p_T^2) d p_T^2 dz \,\sigma^{pp_t\rightarrow f}(s^{\prime}) \equiv d\mathcal{P}_{p \rightarrow p^{\prime}D} \times\sigma^{pp_t\rightarrow f}(s^{\prime}),
\end{align}
\end{widetext}
where the difference in the energy conservation arguments in the delta-functions is neglected in the collinear limit, $p_T\ll p_p$. Here we have introduced the differential splitting probability $d\mathcal{P}_{p \rightarrow p^{\prime}D}=w_{D}(z,p_T^2) d p_T^2 dz$ for radiating a dark state with longitudinal momentum fraction $z$ of the initial beam proton and transverse momentum $p_T$. The splitting functions have the form,
\begin{equation}\label{SplittingScalar}
 w_{S}(z,p_T^2)=\frac{\alpha_\theta}{2\pi}\frac{1}{2H} \bigg[z + z(1{-}z)\bigg(\frac{4m_p^2{-}m_S^2}{H}\bigg) \bigg],
\end{equation}
and 
\begin{align}\label{SplittingVector}
w_{V}(z,p_T^2) &=\frac{\alpha_{\epsilon}}{2\pi}\frac{1}{H}\bigg[z-z(1{-}z)  \bigg(\frac{2m_p^2{+}m_{V}^2}{H}\bigg) {+} \frac{H}{2z m_{V}^2}
\bigg] ,
\end{align}
with $\alpha_{\theta}=g_{SNN}^2\theta^2/4\pi$, $\alpha_{\epsilon}=\alpha_{\rm em}\epsilon^2$. The expression in Eq. (\ref{SplittingScalar}) agrees well with the result of \cite{Boiarska:2019jym} in the case of scalar bremsstrahlung. Adding the time-like and off-shell form factors introduced in~\ref{subsec:FormFactor} to the splitting functions above is straightforward.

\section{WW approximation: \texorpdfstring{$2\rightarrow 2$}{Lg} sub-process}\label{app:2to2}
Consider the 2 to 2 process $p{+}\mathbb{P}\rightarrow p^{\prime}{+}D$, a sub-process of the full 2 to 3 interaction that is relevant for the WW approximation. The matrix elements take the form,
\begin{widetext}
\begin{align}
i\mathcal{M}^{22}_V = ig_VF_{1,V}^p(m_V^2)\tilde{\epsilon}^{\mu}_{q}\epsilon^{\nu\star}_{k}\bar{u}_{p^{\prime}}
\bigg[F_{pp^{\star}D}\big((p{-}k)^2\big)\Gamma_{\mu,\mathbb{P}}(q^2)\frac{i(\cancel{p}{-}\cancel{k}{+}m_p)}{(p{-}k)^2{-}m_p^2}\gamma_{\nu}
{+}\gamma_{\nu}\frac{i(\cancel{p}^{\prime}{+}\cancel{k}{+}m_p)}{(p^{\prime}+k)^2{-}m_p^2}F_{pp^{\star}D}\big((p^{\prime}{+}k)^2\big)\Gamma_{\mu,\mathbb{P}}(q^2)\bigg]u_{p}, 
\end{align}

\begin{align}
i\mathcal{M}^{22}_S = ig_SF_{1,S}(m_S^2)\tilde{\epsilon}^{\mu}_{q}\bar{u}_{p^{\prime}}
 \bigg[\Gamma_{\mu,\mathbb{P}}(q^2)F_{pp^{\star}D}\big((p{-}k)^2\big)\frac{i(\cancel{p}{-}\cancel{k}{+}m_p)}{(p{-}k)^2{-}m_p^2}
 +\frac{i(\cancel{p}^{\prime}{+}\cancel{k}{+}m_p)}{(p^{\prime}+k)^2{-}m_p^2}F_{pp^{\star}D}\big((p^{\prime}{+}k)^2\big)\Gamma_{\mu,\mathbb{P}}(q^2)\bigg]u_{p}, 
\end{align}
\end{widetext}
where $p$, $p^{\prime}$ and $k$ are the momenta of the incoming and outgoing protons and the radiated dark state $D={V,S}$, while $\tilde{\epsilon}^{\mu}_q$ stands for the phenomenological polarization of pomeron with momentum $q^{\mu}$, and the polarization sum is $\sum_{\rm pol}\tilde{\epsilon}^{\mu}_q \tilde{\epsilon}^{\nu}_q{=}{-}g^{\mu\nu}$. We assume that the the on-shell pomeron state is massless, and in the following we set the virtuality $t=q^2$ effectively to zero.

The squared matrix element, averaged (summed) over the initial (final) spins takes the form,
\begin{align}\label{MatEl22}
|\overline{\mathcal{M}_D^{22}}|^2 &= \frac{1}{4}\sum_{\rm spin}|\mathcal{M}_D^{22}|^2 \nonumber\\
 &\approx g_D^2|F_D(m_D^2,m_p^2+U)|^2Y_{\mathbb{P}}^2F_{\mathbb{P}}^2(t)A_D^{22},
\end{align}
where 
\begin{align}
A^{22}_V = 4 \left(2 m_p^2{+}m_V^2\right) \bigg[ & m_p^2 \bigg(\frac{S{+}U}{SU}\bigg)^2 + \frac{S{+}U{-}m_V^2}{SU} \nnl
&  \quad  -2\frac{S^2{+}U^2}{SU}\bigg], 
\end{align}
and
\begin{align}
A^{22}_S = -2 \left(4 m_p^2-m_S^2\right) \bigg[ & m_p^2 \bigg(\frac{S{+}U}{SU}\bigg)^2 + \frac{S{+}U{-}m_S^2}{SU} \nnl
& \quad  -\frac{(S{+}U)^2}{SU}\bigg], 
\end{align}
We have defined the following invariant quantities
\begin{align}
& U \equiv (p-k)^2-m_p^2 = m_D^2-2p\cdot k ,
\\
& S \equiv (p^{\prime}+k)^2-m_p^2 = m_D^2+2p^{\prime}\cdot k .
\end{align}
In the infinite momentum frame, these invariants are related to the kinematic structure function $H(z,p_T^2)$ and take the following simple form
\begin{equation}
U = -\frac{H}{z}, \quad \quad S = \frac{H}{z(1-z)}.
\end{equation} 
Note that we have approximated $F_{pp^{\star}D}(m_p^2{+}S)\approx F_{pp^{\star}D}(m_p^2{+}U)$ in Eq.~(\ref{MatEl22}) in the soft radiation limit.


\bibliography{Refs_pBrem}
\end{document}